\documentclass{aa}
\usepackage[utf8]{inputenc}
\usepackage{multirow}
\usepackage{rotating}
\usepackage{graphicx}
\usepackage{booktabs}
\usepackage{textcomp}
\usepackage{pbox}
\usepackage{natbib}
\usepackage{mathtools}
\usepackage{gensymb}
\usepackage{colortbl}
\usepackage{color}
\usepackage{amssymb}
\usepackage[varg]{txfonts}
\usepackage{verbatim}
\usepackage{scalefnt}
\usepackage[percent]{overpic}
\usepackage{aecompl}
\usepackage{ulem}
\usepackage{hyperref}
\hypersetup{colorlinks,citecolor= [rgb]{0,0,0.85},linkcolor=[rgb]{0,0,0.85}}
\usepackage{todonotes}
\usepackage[english]{babel}
\usepackage{siunitx}
\usepackage{pbox}
\usepackage{float}
\usepackage{lineno}
\usepackage{enumitem}
\usepackage{tikz}
\usepackage{tikz-3dplot}
\usetikzlibrary{angles, quotes}
\usepackage{tkz-euclide}
\usepackage{subcaption}
\usepackage{multicol}
\usepackage[]{threeparttable}
\usepackage{siunitx}
\usepackage{tikz}

\date{March 2023}

\graphicspath{{./}{figures/}}

\begin{document}
\title{Uranus' Complex Internal Structure %\\
%or: \\
%The Challenge in Modeling Uranus' Interior
}

\author{Benno A. Neuenschwander\thanks{Contact e-mail: \href{mailto:benno.neuenschwander@outlook.com}{benno.neuenschwander@outlook.com}} \and Simon M{\"u}ller \and Ravit Helled}
% List of institutions
\institute{Center for Theoretical Astrophysics and Cosmology, Institut f{\"u}r Astrophysik, University of Zurich, \\ Winterthurerstrasse 190, CH-8057 Z{\"u}rich, Switzerland}

\date{Received: 20 September 2023 / Revised: 8 January 2024 / Accepted: 18 January 2024}

\abstract{Uranus' bulk composition remains unknown. Although there are clear indications that Uranus' interior is not fully convective, and therefore has a non-adiabatic temperature profile, many interior models continue to assume an adiabatic interior.}
{In this paper we present a new method to interpret empirical structure models in terms of composition and for identifying non-{ convective} regions. We also explore how the uncertainty in Uranus' rotation period and winds affect the inferred composition and temperature profile.}
{We {  use Uranus' density profiles {from previous work} where the density is represented by up to three polytropes. }  }
{Using our new method, we find that {  these empirical} models {  imply that Uranus' interior includes non-adiabatic regions}. This leads to significantly hotter internal temperatures that can reach a few 10$^3$ K and higher { bulk heavy-element abundances} (up to 1 M$_\oplus$) compared to standard adiabatic models. 
We find that the assumed rotation period strongly affects the inferred composition while the winds have a negligible effect. { Although solutions with only H-He and rock are possible, we find that the maximum water-to-rock ratio in Uranus for our models ranges between 2.6 and 21. This is significantly lower compared to standard adiabatic models.}}
{We conclude that it is {  important} to include non-adiabatic regions in Uranus structure models as they significantly affect the inferred temperature profile, and therefore {  the inferred bulk heavy-element abundance}. In addition, we suggest that it is of great value to measure Uranus' gravitational field and determine its rotation period in order to decrease the uncertainty in Uranus' bulk composition. }

\keywords{Planets and satellites: individual: Uranus - Planets and satellites: interiors - Planets and satellites: composition - Planets and satellites: gaseous planets}

\maketitle

\titlerunning{Uranus' Interior}

\authorrunning{Neuenschwander, M{\"u}ller \& Helled}

\section{Introduction}
Determining the composition, material distribution, and temperature profile within a giant planet is crucial for understanding its formation and evolution. 
Therefore, much effort has been made to investigate and understand the internal structure of Jupiter \citep[e.g.,][]{Hubbard1968,Militzer2016}, Saturn \citep[e.g.,][]{Zharkov_1991, Movshovitz_2020, Militzer_2023}, Uranus and Neptune \citep[e.g.,][]{Podolak_1991,Helled_2011_int_U_N,NETTELMANN2013}. In early work, the interior structure of gaseous planets was assumed to be convective or convective within layers (such as e.g., core, mantle, or envelope). This implies that the entropy throughout the planet/layer is constant and that the temperature profile follows an adiabat which is usually fixed by the planet's known surface properties (typically at 1 bar). 
The reason this method was (and still is) so widespread is its simplicity: for each model with an assumed constant composition of either the entire planet or within its layers, 
the planet's temperature profile and bulk composition are uniquely determined. \\
However, recent studies have indicated that the assumption of a fully convective planet is inappropriate and that all the outer planets in the Solar system show evidence of non-homogeneous regions \citep{2017Wahl,Nettelmann_2021,Miguel_2022,Mankovich_2021,Stanley_2004}. \\
An indication for a non-adiabatic interior for Uranus is its flux. Uranus' flux is nearly in equilibrium with the solar one \citep{Pearl_1990}. Such a low flux, however, is inconsistent with an adiabatic cooling, as it would have taken Uranus much longer than the age of the Solar system to cool down to its present state \citep{Fortney_Nett_2010}. Uranus' low flux could be a result of a thermal boundary that prevents the underlying heat from escaping, leading to a hotter interior \citep{Nettelmann_2016,Scheibe2019,Scheibe2021},  layered convection \citep{Leconte_Chabrier_2013}, composition gradients \cite{Vazan_2020_lumin_U} or a frozen core \cite{Stixrude_2021}. 
Another possibility is that Uranus actually cooled down due to an extreme event like a giant impact, which led to a rapid cooling. This would suggest that Uranus' interior is in fact cold \citep{HF2020}. 
While the origin of Uranus' flux is still unknown, it seems more likely that its low heat flux is a result of a non-adiabatic interior. Three-dimensional numerical dynamo models presented in \cite{Stanley_2004} also indicate the need of a conducting core in both Uranus and Neptune. 
Finally, formation models of Uranus and Neptune imply that the formation process leads to a deep interior with composition gradients that can be sustained for Gyrs \citep[e.g.,][]{Valletta2020,Vazan_2020_lumin_U,Helled2023}. 

Although there are several methods to describe the behavior of non-adiabatic regions \citep[see e.g.,][]{Leconte_Chabrier_2013}, modeling a non-adiabatic planet poses several difficulties. First, it is unclear where the non-adiabatic regions are located and how extended they are. Second, even if the size and extent of a non-adiabatic region are known, its exact characteristics may still be unknown. For example, it is unclear whether this region is fully Schwarzschild stable or allows for double-diffusive convection \citep[see][]{Leconte_Chabrier_2013}. \\
A way to avoid both the simplified assumption of a fully adiabatic planet and the modeling issues of a non-adiabatic planet is to use empirical structure models. In this case the density profile is represented by e.g., a high-order polynomial \citep[e.g.,][]{Helled_2011_int_U_N,Movshovitz_2020}, by polytropes \citep[e.g.,][]{Horedt1983,Neuenschwander_2021} or produced on a random basis \citep[e.g., ][]{Podolak_2022}. Such models determine the density distribution within the planet without a direct link to the physical composition. Empirical structure models can also describe non-standard solutions that include composition gradients and other non-adiabatic regions. \\
By design, empirical structure models do not infer the planetary composition and temperature profile which makes them difficult to interpret and compare to "physical" models (e.g., models based on physical equations of state).

However, given an adequate temperature profile, empirical models can be interpreted in terms of composition. In this study, we present a new method to interpret empirical structure models and determine the planetary temperature profile and  composition using physical equations of state (EoS) of hydrogen, helium water, rock (SiO$_2$) and iron. The method is designed to detect and account for composition gradients and non-{ convective} regions. The inferred  temperature profiles are consistent with the density-pressure profiles determined by  empirical models. 
\\
In this work, we apply this method to various empirical structure models of Uranus and infer its temperature and composition profiles. We then compare these non-adiabatic models to existing adiabatic Uranus solutions and quantify their differences.

This study is structured as follows: In section \ref{sec:method} we explain the newly developed method that we used to infer the temperature and composition profiles of empirical structure models of Uranus. In section \ref{sec:results} we present our findings followed by a summary and conclusion in section \ref{sec:summary_and_conclusion}.

\section{Method} \label{sec:method}
Below we describe how we infer the temperature profile and then interpret the planetary composition of the  density profiles inferred from the empirical structure models presented in  \cite{Benno2022}. 

\subsection{Models: empirical density profiles} \label{subsec:models}
The empirical structure models presented in \cite{Benno2022} are based on up to three polytropes. 
A polytrope relates the pressure $P$ with the density $\rho$:
\begin{equation}
    P = K\rho^{1+ 1/n}, 
\end{equation}
where $K$ (polytropic constant) and $n$ (polytropic index) are free parameters. The three polytropes are arranged piece-wise. This means that each polytropic relation holds only for a specific radial region of the planet. The radii, where the transition from one to the next polytropic relation occurs, are referred to as "transition radii". At these transition radii, density discontinuities can (but don't have to) occur. \\
The free parameters ($K_i$ and $n_i$ (for $i = {1,2,3}$) and transition radii) are chosen such that the resulting density profile fits Uranus' total mass $M$, equatorial radius $a$, rotation rate $\omega$ and gravitational harmonics 
$J_2$ and $J_4$. 
The computation of the gravitational field is achieved with 4$^\text{th}$-order Theory of Figures that calculates the hydrostatic equilibrium (HE) of the planet \cite{Zharkov1970, Zharkov1975, ZharkovVladimirNaumovich1978Popi, Hubbard2014, Nettelmann_2017, Nettelmann_2021}. 
For this calculation we use 4096 radially equally spaced computational layers for the density profiles. To speed up the calculation while maintaining the desired precision, the planet's equipotential shape is calculated every 32$^\text{th}$ layer while using a spline interpolated for intermediate layers. 
More information about e.g., the model calculation method or model parameters can be found in \cite{Neuenschwander_2021, Benno2022}. 

\begin{figure}
    \centering
\includegraphics[width = 0.53\textwidth]{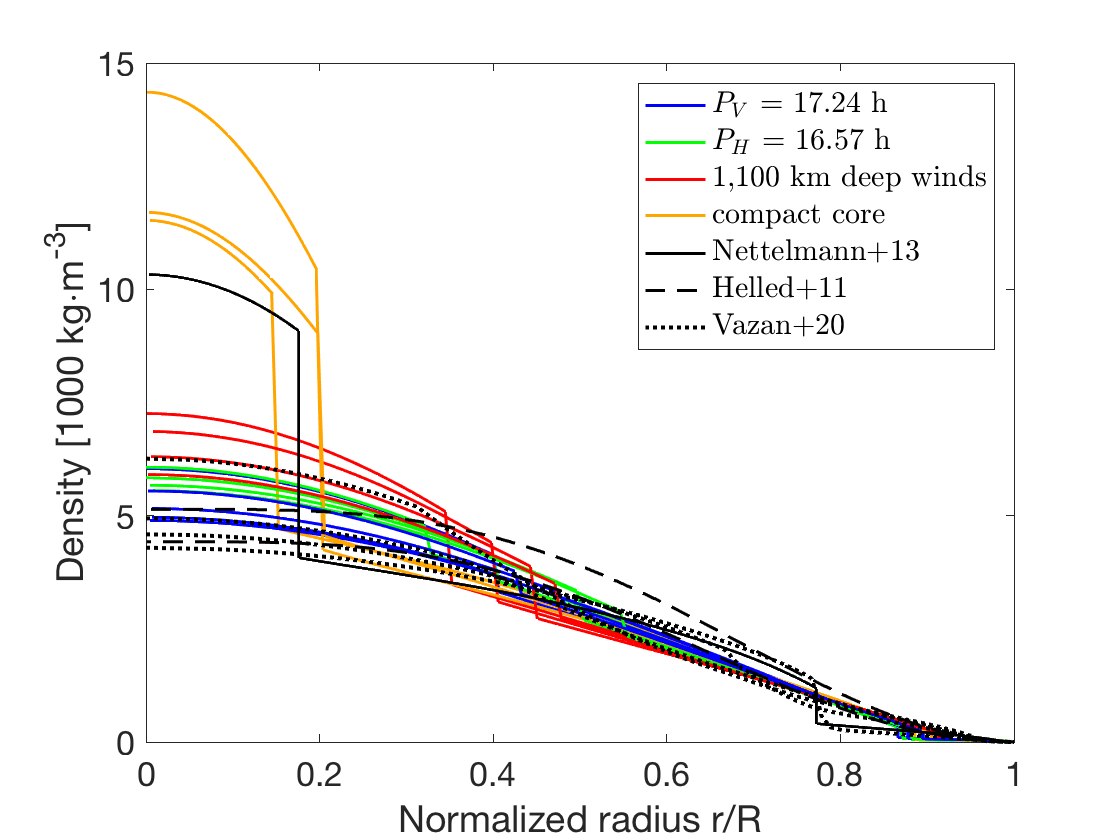}
    \caption{The empirical density profiles{  from \cite{Benno2022} }we consider in this study. Also shown are published density profiles of Uranus. The blue curves show Uranus models that use the rotation period estimated by the Voyager 2 mission, while the green curves show models that use the rotation period proposed by \protect\cite{Helled2010_prot}. The red curves show models with 1,100 km deep winds and the yellow curves correspond to solutions with high central densities. }
    \label{fig:density_profiles}
\end{figure}

Only $J_2$ and $J_4$ have been measured for Uranus, and their values have relatively high uncertainties. In addition, the rotation period of Uranus is not well-determined \citep{Helled2010_uranus_neptune}.  The depth of Uranus' zonal winds also remains unknown,  although depths of $\sim$1,100 km \cite{Kaspi.2013} and $\sim$650 km \cite{Soyuer_2023}) have been estimated. 
Although Uranus' winds are likely to be relatively shallow they still contribute to the measured gravitational field. 
The rotation period of Uranus is not well understood either. 
There are different techniques to estimate the bulk rotation period of Uranus and Neptune: One method is to measure periodicities in the radio signal and the magnetic field of the planet \citep{Desch1986}. A complementary approach is to minimize the dynamical heights on the surface \citep[see][and references therein]{Helled2010_uranus_neptune}. However, these different estimation methods disagree on the order of $\sim$ 40 minutes \citep[see and references therein for more details][]{Helled2010_prot}. 
\cite{Benno2022} showed that both the uncertainty in the depth of Uranus' winds and the large uncertainty in its rotation period affect the internal structure models.  \\ 
We use our newly developed method to investigate the effect of the depth of the winds and the rotation period on Uranus' temperature profile, its bulk composition (including its water-to-rock ratio), and the size and location of non-{ convective} regions. \\
If Uranus' winds are not shallow, we must consider the correction of the winds to the gravitational moments. 
We therefore also present models with the zonal winds penetrating to a depth of 1,100 km. 
We follow \cite{Kaspi.2013} and subtract the expected dynamical contribution from the measured value $J_n^{\text{meas}}$.
In this case, the measured gravitational coefficients $J_{n}^{\text{meas}}$ can be split  into a static $J_{n}^{\text{stat}}$ and a dynamic part $J_{n}^{\text{dyn}}$, where  $J_n^{\text{meas}}$ = $J_{n}^{\text{stat}}$ + $J_n^{\text{dyn}}$. 
In principle, Uranus' winds perturb both its second and fourth gravitational harmonics ($J_2$ and $J_4$). For simplicity, we follow \cite{Kaspi.2013} and only consider the wind effects in $J_{4}^{\text{meas}}$ (i.e.,  $J_{2}^{\text{dyn}}=0$). However, the effect of deep winds on $J_2$ could also affect the density profile and should be considered. We plan to investigate this topic in future research. The empirical models that include the wind contribution are discussed in \cite{Benno2022}.   
Figure \ref{fig:density_profiles} presents the density profiles \citep[taken from][]{Benno2022} that we use in this study. For comparison, we also present other  published density profiles \cite{NETTELMANN2013, Vazan_2020_lumin_U, Helled_2011_int_U_N}. Blue-colored models are models based on the rotation period as measured by the Voyager 2 mission (hereafter $P_{\text{V}} = 17.24$ h) and green models are based on the rotation period as evaluated by \cite{Helled2010_uranus_neptune} (hereafter $P_{\text{H}} = 16.57$ h). Models colored in red account for winds that penetrate to a depth of 1,100 km and, finally, the yellow-colored models represent a "high-density core" (i.e., high central density) solutions. 

As discussed in \cite{Neuenschwander_2021} for Jupiter, the polytropic-based solutions of Uranus published in \cite{Benno2022} tend to overestimate its surface density. As the affected region contains only a tiny fraction of the total planetary mass, its effect on the internal structure model and therefore the published results were negligible. However, since the composition interpretation algorithm is a top-down algorithm as discussed below, it is sensitive to the outer-density profile: An over-dense surface and/or atmospheric region could lead to unrealistically high heavy-element fractions in this region. This, in turn, could prevent the algorithm from converging to solutions below the over-dense region as, by design, the metallicity cannot decrease with depth.  
To avoid this situation, and to obtain more realistic results, the over-dense region has been replaced beforehand with an adiabatic atmosphere model from \cite{Hueso2020}. This substitution only affects the outermost region ($r\gtrsim 0.99$) of Uranus with pressures up to $P \approx 150-250 $ bar. The affected mass is only $\sim 0.002$ M$_\oplus$ which corresponds to $\sim 0.015 \%$ of Uranus' total mass. 
{  \\ Below this region and up to a pressure of $\simeq 10,000$~bar, 
we sometimes encounter situations in the models' convective region wherein the density is lower than expected from a uniform composition. These instances can then lead to violent large-scale convection that would need special treatment. However, the involved mass is less than $1\%$ and the regions with violent large-scale convection only affects the bulk heavy-elements abundance by around $\sim0.01\%$ in mass. Therefore its effect has been neglected in this study. \\
However, it should be acknowledged that empirical models based on three piece-wise arranged polytropes might be insufficient to accurately represent the atmosphere of Uranus. Adding a fourth polytrope to represent the atmosphere can resolve this issue (Morf et al., in prep.). 
}

\subsection{The algorithm to infer the temperature profile} \label{subsec:algorithm}
There are several ways to obtain a suitable temperature profile of an (empirical) model, given its density and pressure profile \citep[see e.g.,][]{Podolak_2019,Podolak_2022}. However, these methods often assume that the planet is fully convective, and apply an adiabatic temperature profile. This assumption may lead to temperature profiles that are not consistent with the composition. If e.g., the empirical density profile predicts a (local) composition gradient, an adiabatic temperature profile fails to describe this region adequately. 

Below, we describe our method to infer the  temperature profile for a given empirical structure model using physical EOS in a self-consistent manner:  
For each model, the method explained in section \ref{subsec:models} yields the density-pressure profile ($\rho(r), P(r)$) but no information about the temperature $T(r)$ or composition profile in the interior. There are several approaches to overcome this issue, which always rely on a physical equation of state. For example, the equation of state could describe $P(\rho, T, \vec{C})$, where $\vec{C} = (X, Y, Z)$ parametrizes the dependence on the composition, and $X$, $Y$, $Z$ are the hydrogen, helium, and heavy-element mass fractions. If the hydrogen-helium ratio is held constant at the proto-solar ratio, then the composition dependence is parametrized by the single value $Z$, since $X + Y + Z = 1$. The problem, however, is that without knowing the temperature, the composition cannot be uniquely determined. This is because the equation $P(r) = P(\rho(r), T(r), Z(r))$ provides a single constraint, but has two unknowns. 

One solution is to assume that the temperature follows a particular adiabat, which means that the temperature profile is known and therefore the composition can be inferred without any ambiguity \citep[e.g.,][]{Helled_2011_int_U_N}. However, these solutions are not self-consistent and can result in adiabatic (and therefore convective) regions with an inhomogeneous  composition. Another approach is to use non-adiabatic temperature gradients. This was done by \citet{Podolak_2019}, who used radiative \citep[e.g.,][]{2013sse..book.....K} and semi-convective \citep{Leconte_Chabrier_2012} temperature gradients to construct non-adiabatic temperature profiles.

Here, we use a different approach: In order to be independent of a prescribed temperature gradient, we have developed a top-down algorithm  (starting from well-defined planetary surface values) that finds { convective} and non-{ convective} regions in the interior. Our equation of state is an updated version that was developed for planetary evolution models \citep{2020AA...638A.121M,2021MNRAS.507.2094M} and combines the \citet{2021ApJ...917....4C} equation of state for hydrogen and helium with the QEoS heavy-element equations of state for H$_2$O, SiO$_2$ and Fe \citep{1988PhFl...31.3059M,2013MNRAS.434.3283V}. 
The algorithm works as follows: At the $P_0 = 1$ bar surface of Uranus, the temperature $T_0$ and density $\rho_0$ are known. We assume that the surface composition only consists of an ideally mixed proto-solar ratio of hydrogen and helium, together with one heavy element. 
Note that we use the ideal-mixing approximation and hence $1 / \rho(P, T) = X/\rho_X(P, T) + Y / \rho_Y(P, T) + Z / \rho_Z(P, T)$, where $\rho_{X, Y, Z}$ are the hydrogen, helium and heavy-element densities at the given pressure and temperature. 
Using $(\rho_0, P_0, T_0$) we can then determine the composition ($X_0$, $Y_0$, $Z_0$) from the EoS. This allows us to also calculate the surface entropy $S_0(P_0, T_0, Z_0)$. Then we move downwards into the planet to the next (computational) layer that is at a higher pressure $P_1$ and density $\rho_1$. Using these two values, we check whether the layer is convective as follows: If the layer is convective, the composition and entropy should stay constant: $Z_1 = Z_0$ and $S_1 = S_0$. Using ($\rho_1$, $P_1$, $Z_1$ = $Z_0$), we calculate the entropy $S_1$ with the equation of state and compare it to the surface value $S_0$. If $S_1 = S_0$, the layer is indeed convective, $T_1$ follows the adiabat, and therefore all required variables ($T_1, Z_1, S_1$) are determined. 

If $S_1 \neq S_0$, the region is not convective and therefore there is potentially a composition gradient such that $Z_1 > Z_0$ (i.e., we force $Z_{i+1} \geq Z_{i}$). In this case, we search for the temperature $T_1$ and composition $Z_1$ that lead to a marginally Ledoux-stable state. In other words, we equalize the Ledoux criterion (see eq. \ref{eq:Ledoux_criterion} below) and solve it to obtain $T_1$ and $Z_1$. The equalized Ledoux criterion \citep{1947ApJ...105..305L} is: 
\begin{equation}
    \nabla_T - \nabla_{ad} = B,
    \label{eq:Ledoux_criterion}
\end{equation}
where $\nabla_T \equiv \frac{\text{d} \ln{T}}{\text{d} \ln{P}}$ is the temperature gradient, $\nabla_{ad} \equiv \left( \frac{\partial \ln{T}}{\partial \ln{P}} \right)_S$ the adiabatic temperature gradient and $B$ is the composition term. $B$ can be written as \cite[see e.g.,][]{Paxton_2013}:
\begin{equation}
    B = -\frac{1}{\chi_T} \frac{\ln{P(\rho_n, T_n, Z_{n+1})} - \ln{P(\rho_n, T_n, Z_{n})}}{\ln{P_{n+1}}-\ln{P_n}},
\end{equation}
where the subscript $n$ denotes the (computational) layer the quantity is assigned to and $\chi_T = \left( \frac{\partial \ln{P}}{\partial \ln{T}} \right)_\rho$. 
{ Note that it is not guaranteed that there is always a ($T_1, Z_1$) tuple that equates the Ledoux criterion. Given, for example, the temperature $T_\text{LD}$ and metallicity $Z_\text{LD}$ that equates equation \eqref{eq:Ledoux_criterion} for layer "1". This solution is rejected if either the temperature or metallicity is smaller than the one of the layer above ($T_\text{LD}<T_0$ and/or $Z_\text{LD}<Z_0$). In this case, out of all the possible  solutions in ($T_1$, $Z_1$) we chose the solution that, first, minimizes $|T_0-T_1|$ then minimizes $|Z_0-Z_1|$ and finally, minimizes $|S_0-S_1|$. In these situations, $T_1$ and $Z_1$ are chosen such that they minimize the change in $T$, $Z$, and $S$ (in this order). 
Again, this yields all required variables ($T_1, Z_1, S_1$) and ensures to minimize the temperature gradient. This approach can lead to regions in the planet that are of positive static stability, i.e., are {stable} against convection. }

Then, the algorithm moves to the next layer, continuing with the same evaluation method described above until the center of the planet is reached. Typically, we analyse the composition of $\sim$1,200 computational layers per model.  
The central temperature is allowed to go up to $T_{max} < 30,000$ K. Note that this generous limit also includes the most extreme models of \citep[see][]{Vazan_2020_lumin_U}. { {  Convective} and non-{ convective} regions are marked as each layer is evaluated and are highlighted/marked in all the figures (where appropriate).} 
By following this procedure we ensure that the composition (and potential composition gradients) are always consistent with the temperature profile. Furthermore, { convective} regions are detected (and accounted for) in an unbiased manner. Note that the detection of non-{ convective} regions depends solely on the combination of pressure, density and the used EoS.

\section{Results} \label{sec:results}
First, we introduce non-adiabatic models that are based on empirical structure models in section \ref{subsec:non_adi models}. We next compare the effects of non-adiabatic regions on the planetary temperature profile and composition with pure adiabatic models in section \ref{subsec:adi_vs_non_adi}. We also explore how the rotation period of Uranus influences the inferred temperature profile and bulk composition in sections \ref{subsec:effect_of_rot_period} and \ref{subsec:water_to_rock_ratio}, respectively. In section \ref{subsec:Wind_effect}, we investigate whether wind dynamics affect Uranus' inferred internal structure. Finally, in section \ref{subsec:compact_core_solutions}, we discuss high-density core solutions of Uranus. 
\\

\begin{figure*}
    \centering
\includegraphics[width = 1\textwidth]{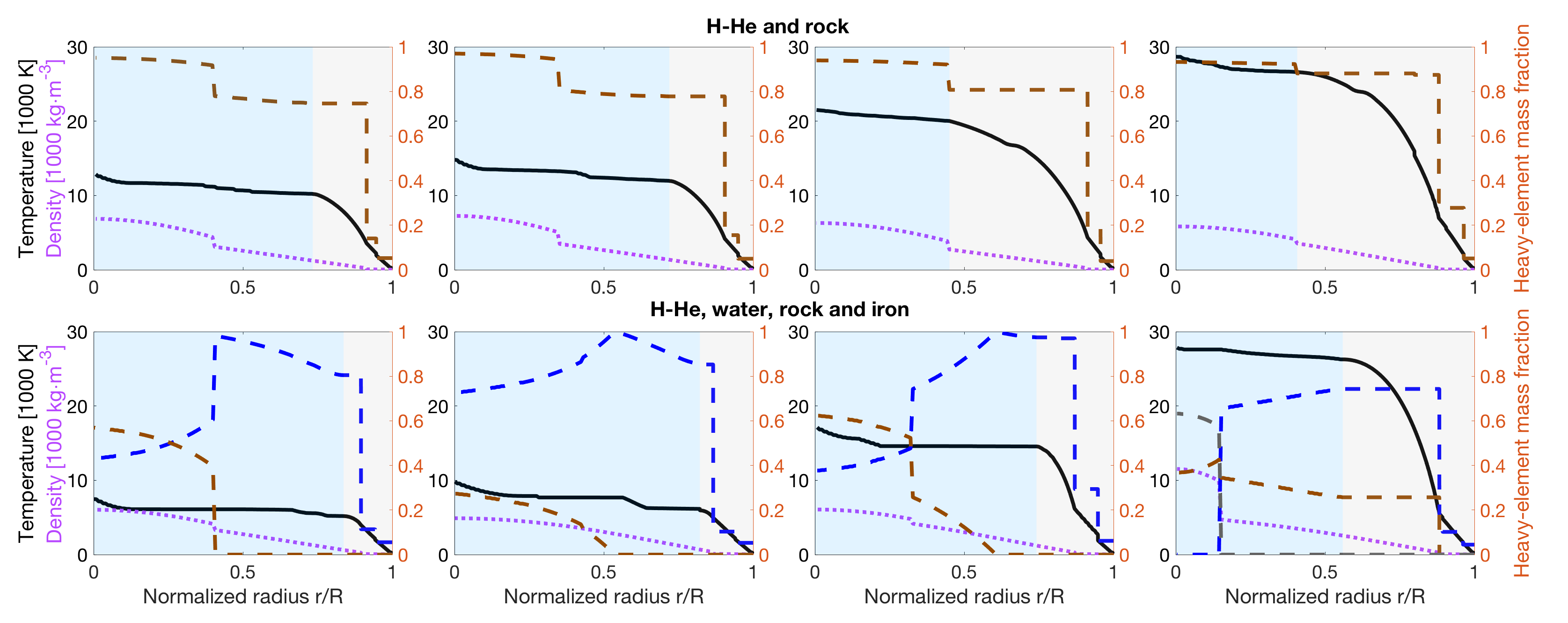}
    \caption{Temperature profiles (solid black curve), heavy-element mass fractions (dashed curves) and density profiles (purple curve) of various solutions.  In the upper row, the heavy elements are represented by pure rock (brown dashed line) only. The lower row shows more "physical" solutions where the heavy elements are represented by water (blue dashed curve), rock (brown dashed curve) and, if necessary, iron (grey dashed curve). The grey-shaded radial area marks the convective regions. On the other hand, the blue-shaded radial area marks the non-{ convective} and non-homogeneous region. The dashed blue, brown, and grey lines show the radial mass fraction of water, rock, and iron, respectively. }
    \label{fig:T_rho_plots}
\end{figure*}

\begin{figure*}
    \centering
\includegraphics[width = 0.93\textwidth]{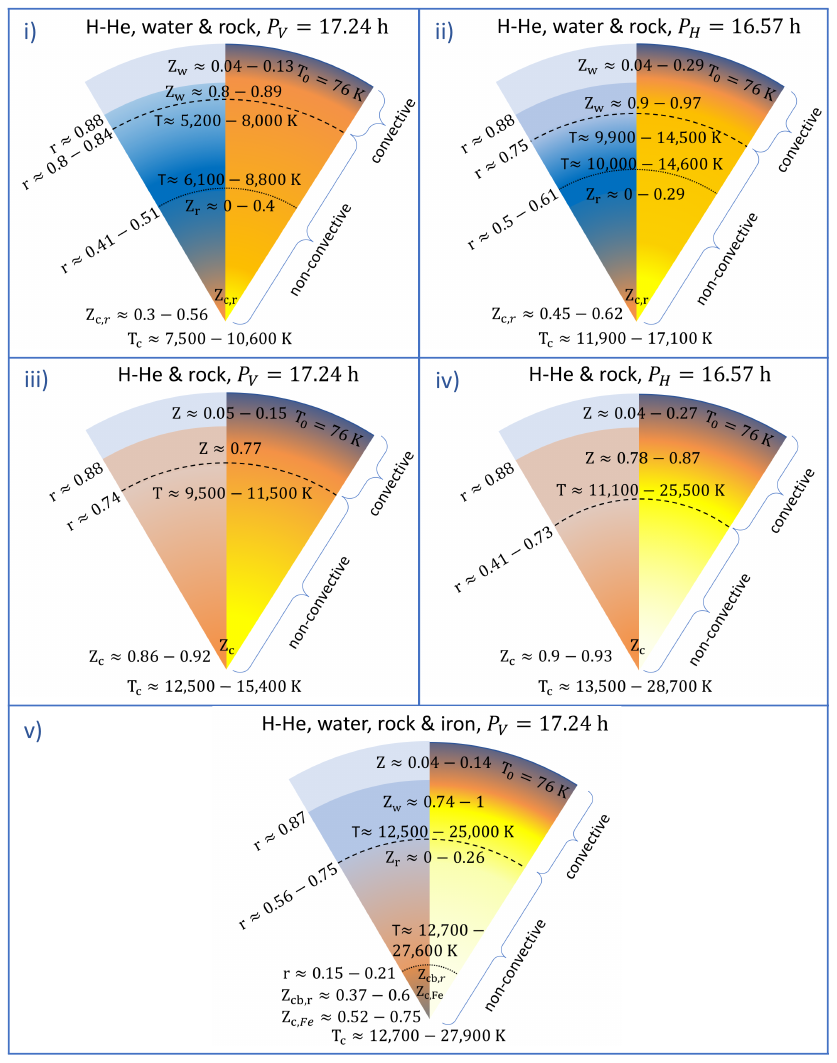}%
    \caption{Sketches of Uranus' interior structure. The left side of each panel shows the composition and the right side illustrates the temperature. Cold regions are colored in blue, hot regions in yellow/white. 
    The protosolar H-He mixture is colored in light blue, water in dark blue, rock in orange and iron (right panel) in grey. Composition gradients are illustrated by a color gradient of the two mixed materials. 
    The dashed curve marks the boundary between the { convective} and non-{ convective} region. Mass fraction of water, rock and iron are indicated with $Z_w$, $Z_r$ and $Z_{\text{Fe}}$. The subscripts 'c' stands for 'central' and 'cb' for 'core boundary'. {  Panel \textbf{i)} represents Uranus with $P_{\text{V}}$ while panel \textbf{ii)} corresponds to Uranus with $P_\text{H}$.
    Panel \textbf{iii)} and \textbf{iv)} illustrate Uranus' interior structure with pure rock as heavy element based on different rotation periods (panel iii):$P_{\text{V}}$, panel iv): $P_{\text{H}}$).
    Panel \textbf{v)} represents Uranus with a high density core.} 
     }
    \label{fig:pie_charts}
\end{figure*}

\subsection{Non-adiabatic Models} \label{subsec:non_adi models}
We analyzed the 16 different density profiles presented in Figure \ref{fig:density_profiles}) and calculate their temperature profile, the material distribution and bulk composition. In addition,  we clearly identify non-adiabatic regions in the planetary interior. \\ 
Figure \ref{fig:T_rho_plots} shows the temperature profiles of a diverse subset covering the diversity of all solutions. It should be noted that in this paper we plot various variables against the normalised radius. To gain an understanding how Uranus' normalized radius compares to its mass, see figure \ref{fig:M_R_relation} in the appendix. The green-shaded radial region in Figure \ref{fig:T_rho_plots} is { convective}, while the blue-shaded radial region is non-{ convective}. The top row shows models  consisting only of a proto-solar mixture of hydrogen and helium, with rock as the heavy element. The density of each model is represented by the purple curve (see also figure \ref{fig:density_figure_2} in the appendix). 
In the bottom row, models are based on a proto-solar H-He mixture together with water, rock, and iron as heavy elements. Note that iron is only needed when the central densities are high (bottom row, most right panel). \\
All the models have non-{ convective} regions covering between 40\% and 80\% of the planet's radius. The exact size  depends on the model, but the inferred structure is always similar: Uranus consists of an { convective} layer that begins at its surface and can extend down to $r\approx 0.4$. Underneath the { convective} region, a non-{ convective} region with a composition gradient extends down to the core. Again, the exact characteristics of the composition gradient are model-dependent. Given the relative size of the non-{ convective} region and its incorporated mass, this region has a non-negligible effect on the planet's temperature profile and bulk composition (see section \ref{subsec:adi_vs_non_adi}). 
We get very diverse solutions in terms of central temperature and metallicity. The core's temperature varies between  7,500 and  28,000 K while the { bulk heavy-element abundance} ranges  between 11.6 and 14.4 (see table \ref{tab:results_rock} \& \ref{tab:results_Z_mixtures} for more details).  

Figure \ref{fig:pie_charts} illustrates the resulting structure, composition and temperature of Uranus based on our non-adiabatic models. The internal structure of Uranus in panel { i)} is based on a mixture of proto-solar H-He, water, and rock. The structure models in panel {  v)} corresponds to high central density models (see section \ref{subsec:compact_core_solutions}) that require iron to be present in the core. Otherwise, it has the same composition as the structure models in panel {  i)}. 
The sketches also show the composition gradient in the non-{ convective} region. Panel {  i)}: Starting at around $r \approx0.8$ the water mass fraction increases up to its maximum around $r \approx0.4 -0.5$. After that, it is successively replaced by rock until it reaches the center of the planet. Panel {  v)}: Starting at around $r \approx0.56 -0.75$ rock increases its mass fraction in the water-rock mixture continuously up to around $r \approx0.15 -0.21$. After this point, water is replaced by iron, which successively increases its mass fraction in the rock-iron mixture down to the planet's center. {  A detailed discussion and comparison of the {  various panels} are provided in section \ref{subsec:water_and_rock}, \ref{subsub:pure_rock} and \ref{subsec:compact_core_solutions}, respectively.}

\subsection{Adiabatic vs Non-adiabatic Interiors} \label{subsec:adi_vs_non_adi}
Here we investigate how non-adiabatic regions within Uranus affect its temperature profile, { bulk heavy-element abundance}, and heavy-element distribution. \\
For this we extracted various adiabatic temperature profiles from published models and, if necessary, linearly extrapolated them to match the required pressures. \\
\begin{figure}
    \centering
\includegraphics[width = 0.52\textwidth]{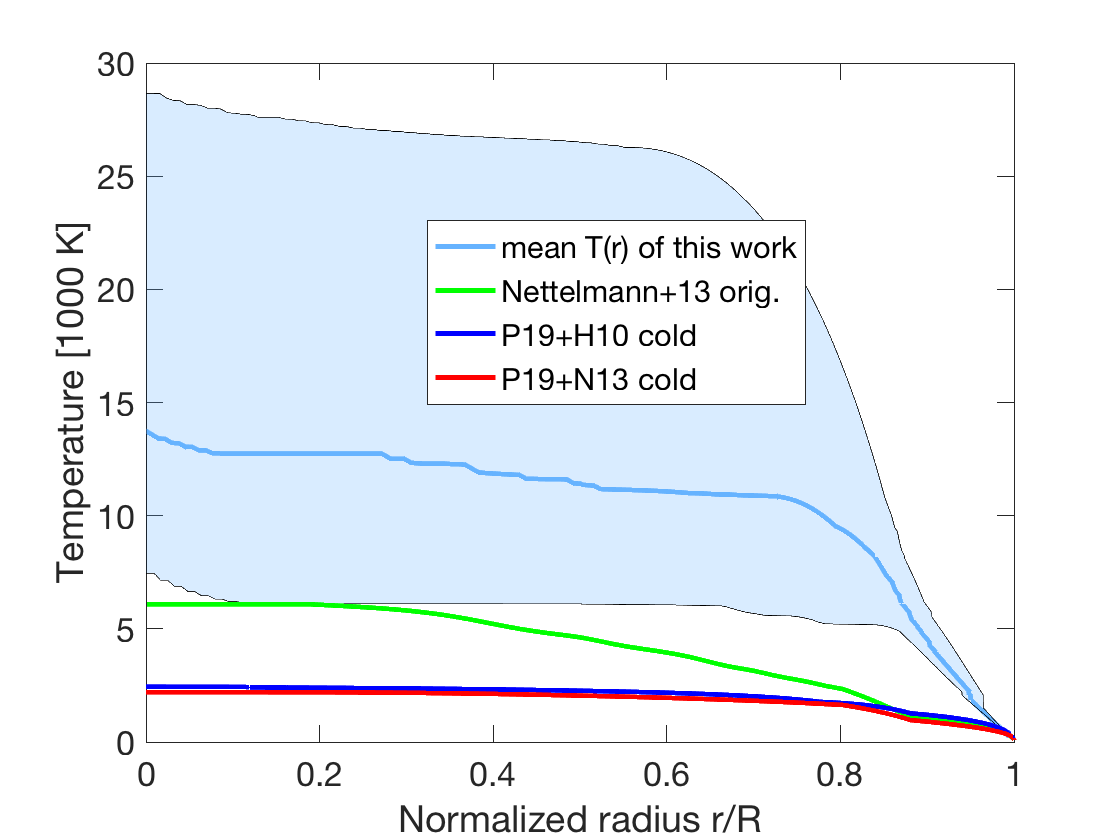}
    \caption{Solution space of temperature profiles of all evaluated models presented in Fig. \ref{fig:density_profiles}. The blue solid curve marks the mean temperature while the blue shaded region marks the 2-sigma uncertainty. For comparison, published adiabatic temperature profiles of \protect\cite{NETTELMANN2013} and \protect\cite{Podolak_2019} are included. }
    \label{fig:Ti_solution_space_poster}
\end{figure}

\begin{figure}
    \centering
\includegraphics[width = 0.52\textwidth]{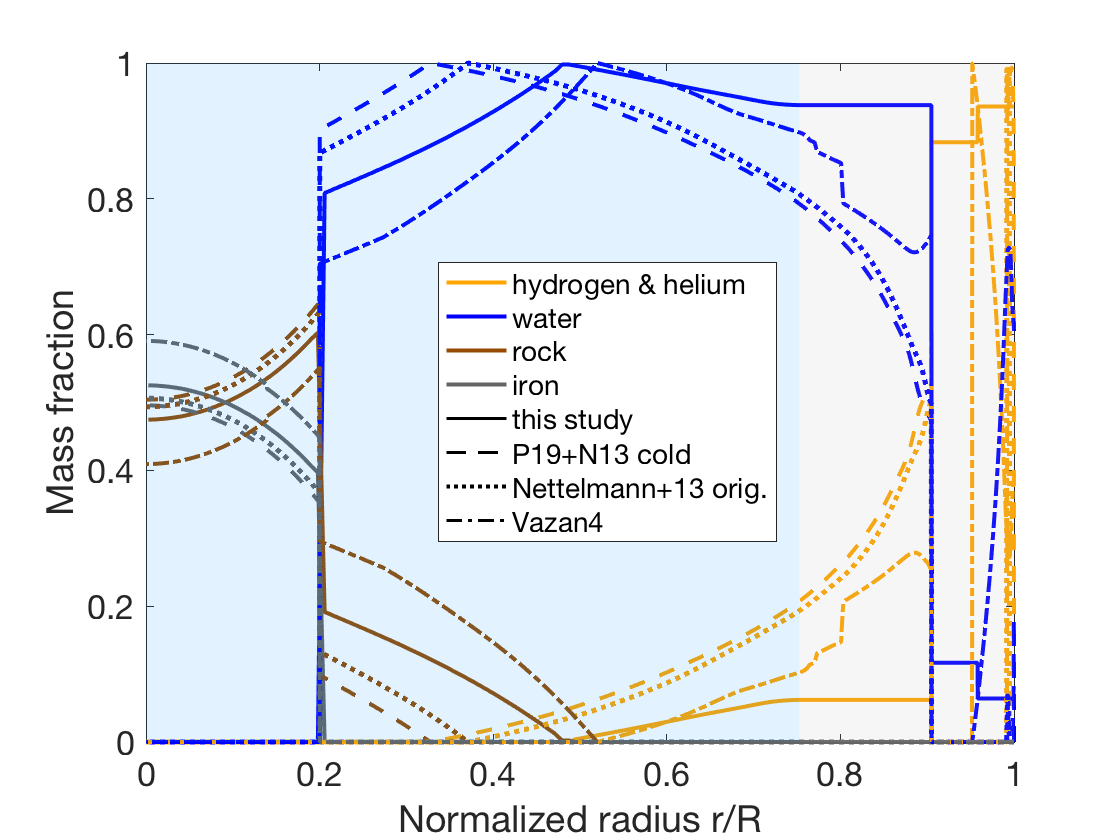}
    \caption{Mass fractions of a dense core model evaluated with different temperature profiles. The yellow curves mark the mass fraction of the proto-solar H-He mixture. The blue, brown and grey curves mark the heavy-element mass fraction of water, rock and iron, respectively. The solid lines mark the mass fractions as evaluated by our method. Mass fractions obtained with the adiabatic temperature profiles "P19+N13 cold" and "Nettelmann+13 orig." are shown with dashed and dotted curves, respectively. Finally, the mass fractions based on the non-adiabatic temperature of "Vazan4" are presented with dash-dotted curves. The grey and blue shaded area indicates the { convective} and non-{ convective} regions, respectively, as identified by our algorithm. 
    }
    \label{fig:compact_core_adi_non_adi}
\end{figure}
Figure \ref{fig:Ti_solution_space_poster} shows the solution space of our non-adiabatic temperature profiles together with published adiabatic temperature profiles. 
We included the adiabatic temperature profiles "U1" from \cite{NETTELMANN2013} (denoted with "Nettelmann+13 orig."), "P19+N13 cold" and "P19+H10 cold" from \cite{Podolak_2019}.
Note that \cite{Podolak_2019} interpreted the  composition of Uranus and Neptune  using the density profiles from \cite{Helled_2011_int_U_N} and \cite{NETTELMANN2013}. For each model, they calculated two temperature profiles, one that follows an adiabat and a second one that accounts for non-adiabatic regions. For the latter they applied the formalism of \cite{Leconte_Chabrier_2012} to obtain a non-adiabatic temperature profile. Both temperature profiles are then used to infer the composition of each model. We refer to the adiabatic models published in \cite{Podolak_2019} as "Podolak+19 cold" and to the non-adiabatic models as "Podolak+19 hot". Accordingly, we refer to their adiabatic and non-adiabatic composition interpretation of models by \cite{NETTELMANN2013} by "P19+N13 cold" and "P19+N13 hot", respectively. The same is true for models of \cite{Helled_2011_int_U_N}: "P19+H10 cold" and "P19+H10 hot". 
Figure \ref{fig:Ti_solution_space_poster} clearly shows that non-adiabatic models have significantly hotter interiors compared to adiabatic models. { This is expected as most of Uranus' interior is {stable} and heat dissipates much slower in these non-convective regions which then trap primordial heat and stay hot for a longer time, leading to hotter present-day interiors.} \\
We use the external adiabatic temperature profiles described above together with the non-adiabatic model model "U-4" (denoted as "Vazan4") from \cite{Vazan_2020_lumin_U} to infer the composition of an arbitrary dense core model using a protosolar H-He mixture, together with water, rock and iron as heavy elements. For a simpler comparison with only rock as heavy element, see appendix \ref{app:adi_vs_nonadi_pure_rock}. We also applied our method to infer the non-adiabatic temperature profile and composition of the same model. This allows us to compare the inferred composition of adiabatic and non-adiabatic temperature profiles against each other. \\
Figure \ref{fig:compact_core_adi_non_adi} shows an example of the inferred distribution of H-He, water, rock and iron using our method (solid curves). For comparison, we also present the composition distributions evaluated with published adiabatic and non-adiabatic temperature profiles ("P19+N13 cold" and "Nettelmann+13"). 
Note, however, that these external temperature profiles are not consistent with the empirical $P(r)$ and $\rho(r)$ from the model and therefore could lead to non-physical solutions. Since we mostly observe  a non-physical behavior in the outermost regions (upper $\sim10\%$ in radius), these regions are excluded. 
Excluding the upper most $\sim10\%$ does not change the results to any significant extent, since the mass associate is only about $\sim1\%$ of the planet's mass.\\ 
The adiabatic structure model assumes that the composition is homogeneously mixed throughout the planet. However, the adiabatic temperature profiles ("P19+N13 cold" and "Nettelmann+13 orig.") suggest that there is a composition gradient, meaning that the composition changes with depth. This is a contradiction, and it shows that neither of the adiabatic temperature profiles agrees with the structure model. \\
We find that different adiabatic temperature profiles have little impact on the distribution of the heavy elements and the { bulk heavy-elements abundance}. For "P19+N13 cold", we obtain $Z_{\text{H}2\text{O}} = 11.90 \text{ M}_\oplus$, $Z_{\text{SiO}2} = 0.60 \text{ M}_\oplus$ and $Z_{\text{Fe}} = 0.38 \text{ M}_\oplus$, while for "Nettelmann+13", we obtain $Z_{\text{H}2\text{O}} =12.00 \text{ M}_\oplus$, $Z_{\text{SiO}2} = 0.65 \text{ M}_\oplus$ and $Z_{\text{Fe}} = 0.40 \text{ M}_\oplus$. The differences are around 0.1 M$_\oplus$. However, when we compare the adiabatic "P19+N13 cold" to the non-adiabatic "Vazan4" ($Z_{\text{H}2\text{O}} = 12.02 \text{ M}_\oplus$, $Z_{\text{SiO}2} = 1.19 \text{ M}_\oplus$ and $Z_{\text{Fe}} = 0.47 \text{ M}_\oplus$), we see larger differences of up to $\sim 0.8$ M$_\oplus$. This result is expected because in general, the density of material decreases with increasing temperature. Therefore, a hot region with a given density can store more heavy elements than the same region with a cooler temperature. \\
Uranus' deep interior is expected to be composed of refractory materials where the density is less sensitive to temperature. Therefore, although the inferred composition is not expected to be very sensitive to an increase in temperature, we find a difference of 0.5 M$_\oplus$ for the inferred { bulk heavy-element abundance} for the non-adiabatic models, and a difference in the rock-to-water ratio of the order of 50\%,  compared to non-adiabatic models. 
While \cite{Nettelmann_2017} inferred for an  adiabatic model a water-to-rock ratio of 19-35, our non-adiabatic models predict a water-to-rock ratio between 7 and 15 (for more information see section \ref{subsec:water_to_rock_ratio}). \\
We conclude that different adiabatic temperatures do not alter the resulting { bulk heavy-element abundance} in a significant way.
However, the hotter non-adiabatic temperatures can significantly affect the { bulk heavy-element abundance} and definitely need to be accounted for. The non-adiabatic regions also significantly affect the planetary long-term evolution \citep{Vazan_2020_lumin_U,Scheibe2019,Scheibe2021}. Therefore, a precise description of the non-adiabatic regions within Uranus is key to constrain its formation and evolution path. For these reasons, it is crucial that internal structure models of Uranus take into account non-convective regions.

\begin{figure}
    \centering
\includegraphics[width = 0.25\textwidth]{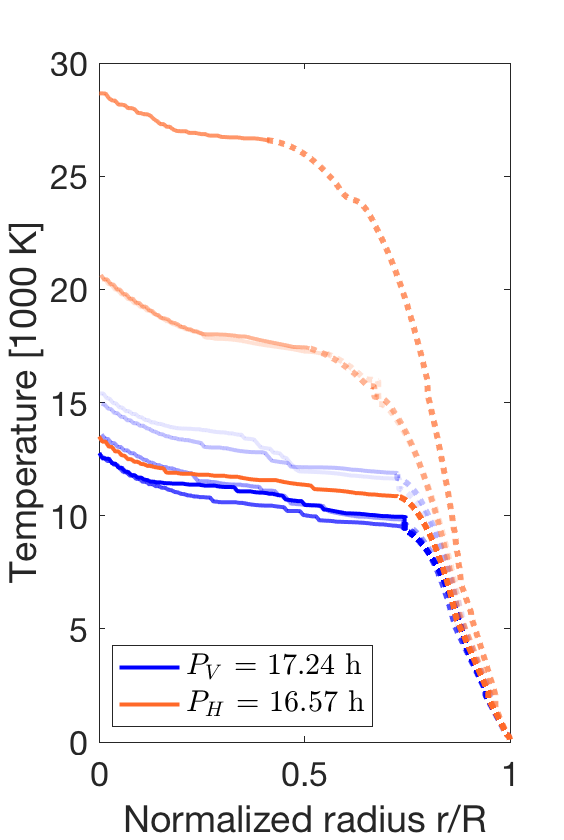}%
\includegraphics[width = 0.25\textwidth]{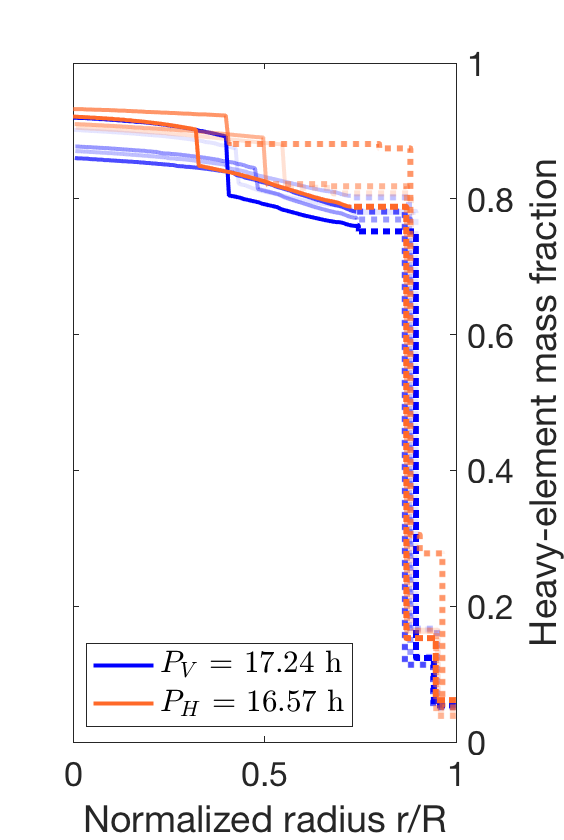}%
    \caption{Temperature (left panel) and metallicity (right panel) profiles of Uranus models with different rotation periods. Different orange and blue shades are used to make individual models traceable across the two panels. The dotted (solid) part of each curve corresponds to { convective} (non-{ convective}) regions. }
    \label{fig:diff_rot_T_Z}
\end{figure}

\subsection{Effect of the rotation period on the composition} \label{subsec:effect_of_rot_period}

In \cite{Benno2022} (see section 3.2.1) we explored the sensitivity of the inferred density profile to the assumed planetary rotation period. It was concluded that the mean central density for a faster rotating Uranus (with $P_{\text{H}}$) is 13\% higher than for a Uranus with $P_{\text{V}}$. \\
In this section we investigate the effect of different rotation periods on the temperature profile, the distribution and total amount of heavy elements, and the water-to-rock ratio.
We chose a subset of models presented in \cite{Benno2022}. Although this subset does not cover the entire solution space (shown in Figure 2, bottom left panel, of \cite{Benno2022}), it still represents its mean density.

\subsubsection{Pure-rock}   \label{subsub:pure_rock}
To explore the effect of different rotation periods on the inferred composition and temperature in isolation, we first interpret Uranus' models in the simplest way where the heavy elements are represented solely by rock, together with H-He at a proto-solar ratio. This oversimplified assumed composition is clearly unrealistic but has the advantage that no assumptions on the ratios between various heavy elements need to be made. This assumption allows us to isolate the effect of the assumed rotation rate on the inferred composition. In the next subsection, a more realistic model is used. 
 Figure \ref{fig:diff_rot_T_Z} shows the temperature profile (left panel) and heavy-element distribution (right panel) of Uranus models using  $P_{\text{H}}$ (orange-themed) and $P_{\text{V}}$ (blue-themed).
We find that models with $P_{\text{H}}$ can exhibit significantly hotter interiors that are more enriched with heavy elements.  This result is  in agreement with \cite{NETTELMANN2013}. 
Core temperatures over 15,000 K are only found in a Uranus with $P_{\text{H}}$. Further, {models based on $P_{\text{H}}$ have} an increased { bulk heavy-elements abundance} of around $\sim5\%$ (or 0.6 M$_\oplus$). As shown in Table \ref{tab:results_rock} the { bulk heavy-elements abundance} increases from 11.6-11.9 M$_\oplus$ (for $P_{\text{V}}$) to 11.9-12.2 M$_\oplus$ (for $P_{\text{H}}$). 
Finally, the rotation period affects the size of the { convective} region within the planet:  for  Uranus models with  $P_{\text{V}}$, $\sim$ 25\% of the planet is { convective}, while models with $P_{\text{H}}$ have { convective} regions that  can cover up to $\sim$ 60\% of the planetary radius. The size of the { convective} region within Uranus can lead to a significantly different cooling history \citep{Scheibe2021}. Therefore, an accurate measurement of the rotation period of Uranus is key to constrain its formation and evolution path. \\
Figure \ref{fig:pie_charts} { (panel iii) and iv))} shows sketches of Uranus' structure and temperature for the two different rotation periods. Although qualitatively similar, there are differences in the heavy-element mass fraction and the size of the { convective} region. \\

\subsubsection{Water and rock} \label{subsec:water_and_rock}

To investigate the effect of the rotation period on a more realistic configuration, we allow Uranus to consist of H-He, water, and rock.  
We interpret the same models as in the previous section. Since the internal structure of Uranus is now more complex, the bulk  composition is derived as follows: \\ 
We assume that Uranus' atmosphere consists of a mixture of H-He and water. In the non-convective region, the water mass fraction begins to steadily increase towards the center until it reaches $Z_w = 1$ (and hence $X = Y= 0$). The H-He mixture is then replaced by rock which increases (in the presence of a composition gradient) towards the planetary center, while the water fraction decreases (to assure ($Z_{total} = Z_w + Z_r = 1$). \\
\begin{figure*}
    \centering
\includegraphics[width = 0.5\textwidth]{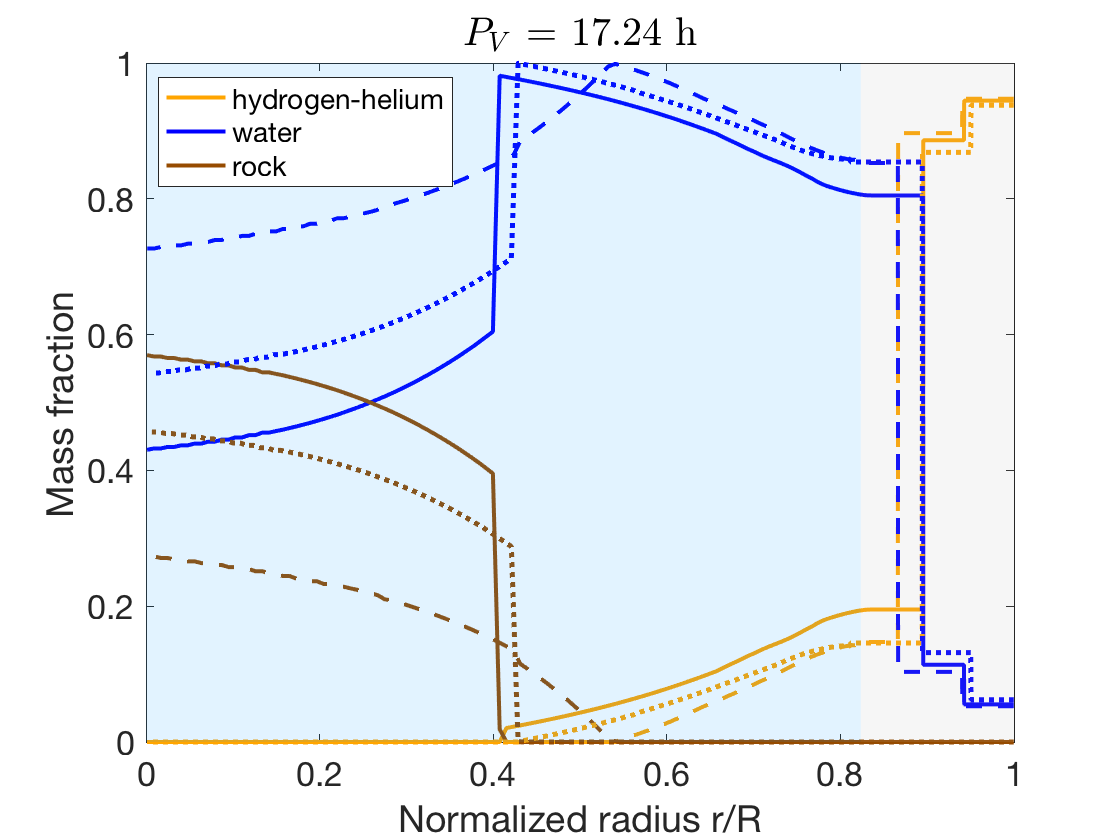}%
\includegraphics[width = 0.5\textwidth]{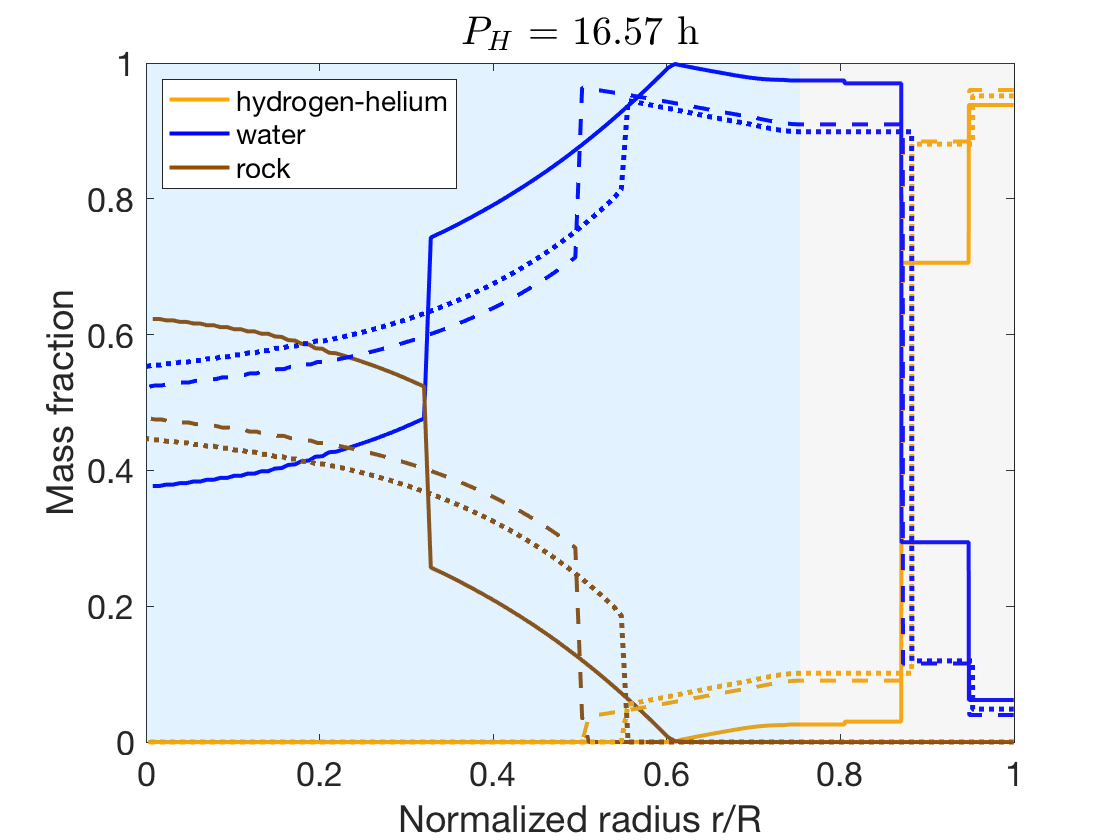}%
    \caption{The radial mass fractions of the H-He mixture, water and rock of Uranus models using $P_{\text{V}}$ (left panel) and $P_{\text{H}}$ (right panel). The solid, dashed, and dotted curves show the mass fraction of different models. In the left panel, the solid and dashed curves show the most extreme models in terms of rock and water mass fractions, whereas the dotted curves mark a model in-between. The grey and blue-colored areas mark in average the { convective} and non-{ convective} regions of the presented models. }
    \label{fig:diff_rot_Zi_mix}
\end{figure*}
Figure \ref{fig:diff_rot_Zi_mix} shows the distribution of the H-He, water, and rock mass fraction for a selection of models with $P_{\text{V}}$ (left panel) and $P_{\text{H}}$ (right panel). The main findings are similar: We find that models with $P_{\text{V}}$ have higher { bulk heavy-element abundances} and higher mass fractions of denser material (i.e., rock). This in turn affects the overall water-to-rock ratio (see next section \ref{subsec:water_to_rock_ratio}). Again,   models with $P_{\text{V}}$ have a larger { convective} region in Uranus' atmosphere/mantle (by 35\%). Finally, Uranus' interiors with a core temperature $\gtrsim 10,000$ K can are found for Uranus models with the faster rotation period ($P_{\text{H}}$).  \\
The inclusion of water as a heavy-element affects the inferred bulk composition of Uranus. The H-He { bulk abundance} is roughly halved (below $\sim$1.25 M$_\oplus$) in comparison to the pure rock solution ($\sim$2-3 M$_\oplus$, see table \ref{tab:results_rock} and \ref{tab:results_Z_mixtures}). This is because, in the pure rock scenario, the H-He mixture mixes with rock down to the planetary center. In the case where the heavy elements are represented by water and rock, the H-He mixture gets substituted relatively quickly by water. \\

Uranus models with $P_{\text{H}}$ can exhibit  higher central temperatures, which is a consequence of the higher { bulk heavy-elements abundance} (and lower water-to-rock ratio) (see table \ref{tab:results_Z_mixtures}). \\
Figure \ref{fig:diff_rot_Zi_mix} shows that similar to the pure-rock solutions, the { convective} region is larger for the faster-rotating Uranus with $P_{\text{H}}$. 
However, in comparison with the pure rock solutions, the transition radius between the { convective} atmosphere and the non-{ convective} deeper interior is, independent of the rotation period, further out. This indicates that a more physical interior of Uranus is largely non-convective. This result has important implications for Uranus evolution models.

Figure \ref{fig:pie_charts} { (panel i) and ii))} presents Uranus' internal composition and temperature when both water and rock are included. Note that panel { i) and ii)} represent a Uranus with $P_{\text{V}}$ and $P_{\text{H}}$, respectively. As for pure-rock, the { convective} atmosphere is dominated by H-He. This is followed by a composition change around $r\approx0.88$ where water becomes the dominant constituent (at least in terms of mass). In the non-{ convective} region, the water mass fraction steadily increases until a pure-water\footnote{Note that a pure-water layer is unlikely to exist in reality since pollution of other refractory materials is expected.} region is reached. After this point, the rock mass fraction keeps increasing towards the planetary center.
Interestingly, the transition to a rock-rich layer occurs further out for Uranus models with $P_{\text{H}}$.  
It should be noted that recent studies suggest that water and rock could be miscible in Uranus' deep interior  \citep{Kovacevic_2022,Pan_2023}, see also section \ref{subsec:water_to_rock_ratio}. If this is correct, the deep interior of Uranus is expected to be even more rock-rich. It is therefore clear that improved knowledge of rock-water mixtures is required.

\subsubsection{High central densities and the inclusion of iron} \label{subsec:compact_core_solutions}
\begin{figure}
    \centering
\includegraphics[width = 0.52\textwidth]{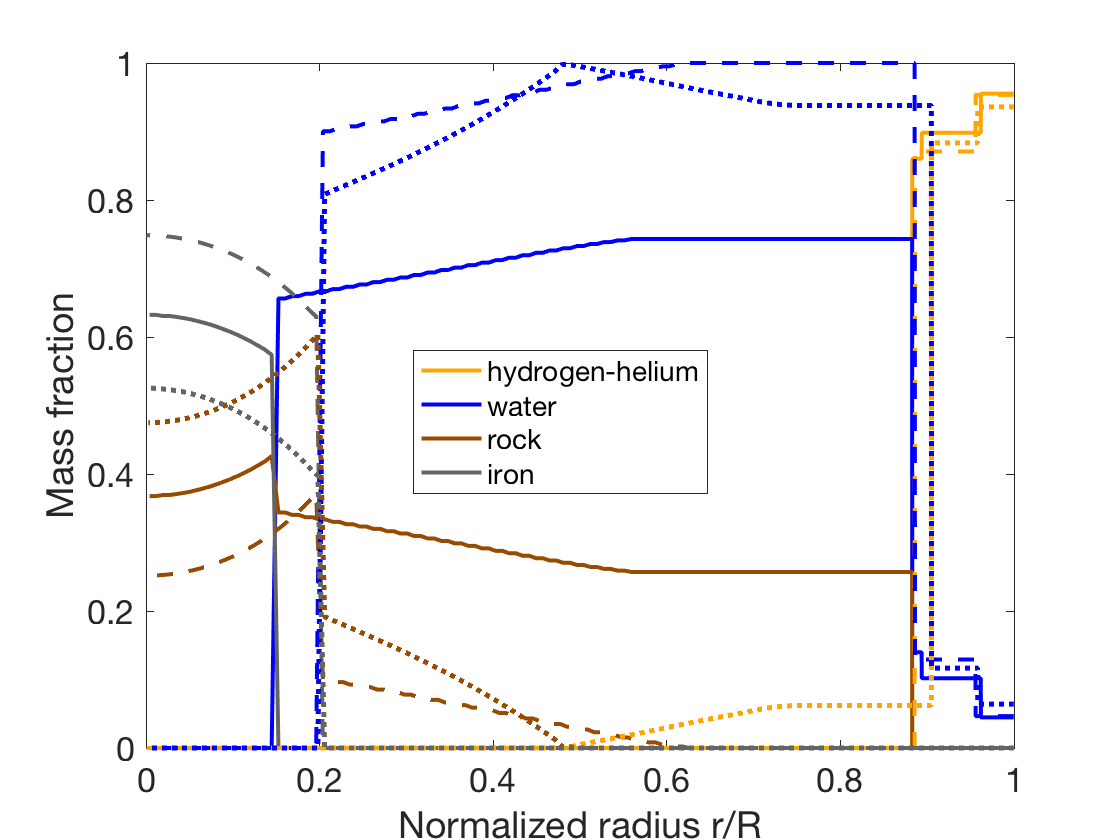}
    \caption{Mass fractions of three high central density models. The different models are represented by the different lines (solid, dashed, and dotted) }
    \label{fig:compact_core_Zi}
\end{figure}

So far, our Uranus models only consist of H-He where the heavy elements are represented by water and rock. This mixture is more appropriate for models where the central densities are relatively low ($<\rho_{core} \approx7000 \text{ kg/m}^3$). For models with high central densities  (see orange curves in figure \ref{fig:density_profiles}) Uranus' central density exceeds the density of pure rock under the inferred pressure and (calculated) temperature. For these models, the inclusion of a denser material is needed, for which we use iron. 
Note that in Uranus, iron is only needed for densities above $\sim 7,000-8,500$ kg/m$^3$ where the exact number depends on the actual pressure and temperature.  \\
Figure \ref{fig:compact_core_Zi} shows the distribution of H-He, water, rock, and iron for three Uranus models with high central densities. 
Note that iron replaces water at the point where the water-rock mixture is no longer dense enough to match the planetary density profile.
The inferred { bulk abundance} of iron in these models is rather small and is between 0.22-0.7 M$_\oplus$. 
In this analysis, high central density models have the lowest water-to-rock ratio (as low as 2.6) and exhibit the lowest central entropy. \\
Figure \ref{fig:pie_charts} (panel v)) shows the structure of a high central density model. It consists of a convective and metal-poor atmosphere above a convective water-rich layer. Below the convective region, water is gradually replaced by rock down to the core-envelope boundary. The non-convective "core" (innermost region) consists of rock and iron, where the mass fraction of iron is steadily increasing up to the planetary center. The transition into the non-{ convective} region occurs deeper in the planet compared to solutions with only water and rock as heavy elements. 
This effect, however, is not necessarily a direct consequence of including iron as a heavy element. Rather, it arises directly from the density profile, which is different for models with a high central density. Since more mass is concentrated in the innermost region, the mantle and/or envelope of the planet are less massive. This leads to a flatter density gradient in the mantle, which, at least partially, does not have to be described by a material gradient.

\subsection{Water-to-rock ratio}  \label{subsec:water_to_rock_ratio}
Knowledge about the water-to-rock ratio in Uranus is crucial for constraining Uranus' formation history.  
The water-to-rock ratio of Uranus (and Neptune) is unknown. Since Uranus has probably formed outside the water-ice-line, condensed water ice is expected to be abundant. On the other hand objects near Uranus (and Neptune), e.g., Kuiper belt objects, are found to be rock-rich \citep[e.g.,][]{Malamud_2015, Bierson_2019}. Also, the Pluto system is rock-dominated \citep[][and references therein]{Stern_2018}, as well as other Kuiper belt dwarf planets and comets \citep[][and references therein]{Podolak_2022}. As a result, Uranus might also be rock-dominated in composition \citep[e.g.,][]{Teanby2020}. 
Although it has been shown that Uranus' measured properties can be reproduced even with an internal structure without any water \citep[e.g., ][]{Helled_2011_int_U_N}, typically, structure models predict that Uranus is water-dominated in composition. For example, \cite{NETTELMANN2013} infer a water-to-rock ratio of 19-35 for Uranus. \\ 
Below, we determine the water-to-rock ratios from our models and show that although our method prefers water over rock, the inferred water-to-rock ratio can be significantly lower compared to previous studies.  
In the section above, we showed that the water-to-rock ratio of Uranus depends on its rotation period. For Uranus models with  $P_{\text{V}}$, the water-to-rock ratio is between 7 and 15, whereas for Uranus models with $P_{\text{H}}$ it is between 5 and 6. Due to the higher metallicity of the models with $P_{\text{H}}$,  the transition to rock starts further out in the planet (around $r\approx 0.5-0.61$ instead of $r\approx 0.41-0.51$). This increases (lowers) the total amount of rock (water) and leads to a decreased water-to-rock ratio. 
It should be noted, however, that these water-to-rock ratios mark an upper bound. The way the planetary composition is inferred is biased towards a high water content because the heavy elements are represented exclusively by water until the model density exceeds the density of pure water, and only at that point rock is added.  
In reality, both water and rock are expected to be mixed in the outer regions of Uranus, which would decrease the water-to-rock ratio.   Previous studies support that scenario by demonstrating that water and rock are miscible in conditions that exist in Uranus \citep[e.g.,][]{Kim2021,Soubiran2017,Pan_2023}. Finally, other molecules (such as CO or MgO) could be present that might perturb the overall water-to-rock ratio. \\
 Finally, it is also important to keep in mind that we can represent Uranus' interior without the presence of water at all, and its density profile can be reproduced by a mixture of H-He and rock (see section \ref{subsub:pure_rock}). {  Note that in the compact core solution, in principle, water, rock and iron could co-exist in the core. This is however, not very likely, and also corresponds to a very small mass. Therefore, our estimates to the water-to-rock ratios could be taken as upper bounds. } 
\begin{table*}
\centering
\begin{threeparttable}
	\caption{Key parameters for all evaluated models consisting of a mixture of H-He and rock.}
	\begin{tabular}{lccccccc}
    \textit{model name} & $T_c$ [K]  & $Z_c$    & $S_{0} - S_c$ [erg/K/g]  & $r_{\text{a}\rightarrow \text{na}}$ & $XY_{\text{tot}}$ [M$_\oplus$] & $Z_{\text{tot}}$ [M$_\oplus$] & $\rho_{c}$ [kg/m$^3$] \\
    \hline
    model 1 $P_{\text{V}}$     & 12,768   & 0.919  & 8.598 - 7.772   & 0.743    & 2.94   & 11.59   & 6035   \\
    model 2 $P_{\text{V}}$     & 12,562   & 0.860  & 8.599 - 7.963   & 0.739    & 2.89   & 11.64   & 4893   \\
    model 3 $P_{\text{V}}$     & 13,591   & 0.877  & 8.598 - 7.924   & 0.741    & 2.88   & 11.65   & 5159   \\
    model 4 $P_{\text{V}}$     & 14,943   & 0.871  & 8.596 - 7.956   & 0.725    & 2.64   & 11.89   & 4964   \\
    model 5 $P_{\text{V}}$     & 15,399   & 0.902  & 8.596 - 7.865   & 0.730    & 2.62   & 11.82   & 5546   \\
    \hline
    model 1 $P_{\text{H}}$     & 13,501   & 0.921  & 8.596 - 7.773   & 0.727    & 2.68   & 11.85   & 6070   \\
    model 2 $P_{\text{H}}$     & 28,658   & 0.932  & 8.599 - 7.828   & 0.407    & 1.66   & 12.87   & 5834   \\
    model 3 $P_{\text{H}}$     & 20,638   & 0.910  & 8.603 - 7.869   & 0.504    & 2.30   & 12.23   & 5672   \\
    model 4 $P_{\text{H}}$     & 20,427   & 0.905  & 8.600 - 7.887   & 0.554    & 2.29   & 12.24   & 5556   \\
    \hline
    model 1 winds              & 14,794   & 0.970  & 8.599 - 7.527   & 0.720    & 2.75   & 11.79   & 7254   \\
    model 2 winds              & 12,855   & 0.951  & 8.599 - 7.624   & 0.734    & 2.95   & 11.59   & 6859   \\
    model 3 winds              & 21,509   & 0.939  & 8.603 - 7.756   & 0.451    & 2.30   & 12.24   & 6302   \\
    model 4 winds              & 15,512   & 0.915  & 8.599 - 7.813   & 0.721    & 2.76   & 11.78   & 5907   \\
    \end{tabular}
	\begin{tablenotes}
	    \item The models are separated into three groups: The first group uses the slower rotation period of Uranus ($P_{\text{V}}$) to calculate its structure. The second group uses the faster rotation period ($P_{\text{H}}$) instead. The third group also uses $P_{\text{V}}$, but adds a wind layer that is 1,100 km deep. $T_c$ denotes the central temperature and $Z_c$ the central metallicity (mass fraction of rock). $S_{0}$ and $S_c$ denote the entropy on the surface and in Uranus' center. $r_{\text{a}\rightarrow \text{na}}$ is the radius at which Uranus becomes non-{ convective}. $XY_{\text{tot}}$ and $Z_{\text{tot}}$ are the total bulk abundances of the H-He mixture and rock, respectively. Finally, $\rho_{c}$ is the central density.
    \end{tablenotes}
	\label{tab:results_rock}
\end{threeparttable}
\end{table*}

\begin{table*}
\centering
\begin{threeparttable}
    \caption{Key parameters for all evaluated models consisting of H-He, water, rock and potentially iron.}
    \begin{tabular}{lccccccccc}
    model name & $T_c$ {[}K{]} & $S_0-S_c$ [erg/K/g]  & $r_{\text{a}\rightarrow \text{na}}$ & $XY$ {[}M$_\oplus${]} & $Z_{\text{H}_2\text{O}}$ {[}M$_\oplus${]} & $Z_{\text{SiO}_2}$ {[}M$_\oplus${]} & $Z_{\text{Fe}}$ {[}M$_\oplus${]} & w:r ratio & $\rho_{\text{c}}$ {[}kg/m$^3${]} \\
    \hline
    model 1 CC  & 27,854  & 8.601 - 7.339  & 0.559  & 0.12  & 10.24  & 3.96  & 0.22  & 2.6   & 11523  \\
    model 2 CC  & 20,182  & 8.605 - 7.275  & 0.626  & 0.14  & 13.07  & 0.63  & 0.70  & 20.7  & 14358  \\
    model 3 CC  & 12,740  & 8.595 - 7.201  & 0.753  & 0.48  & 12.78  & 0.87  & 0.40  & 14.7  & 11701  \\
    \hline
    model 1 $P_{\text{V}}$  & 7,457   & 8.598 - 7.334  & 0.837  & 1.25  & 11.60  & 1.68  & 0     & 6.9   & 6035   \\
    model 2 $P_{\text{V}}$  & 9,822   & 8.599 - 7.505  & 0.822  & 0.81  & 12.88  & 0.84  & 0     & 15.3  & 4893   \\
    model 3 $P_{\text{V}}$  & 10,600  & 8.598 - 7.489  & 0.826  & 0.96  & 12.33  & 1.25  & 0     & 9.9   & 5159   \\
    model 4 $P_{\text{V}}$  & 9,773   & 8.596 - 7.496  & 0.801  & 0.66  & 12.97  & 0.91  & 0     & 14.2  & 4964   \\
    model 5 $P_{\text{V}}$  & 9,135   & 8.596 - 7.419  & 0.812  & 0.92  & 12.24  & 1.38  & 0     & 8.9   & 5546   \\
    \hline
    model 1 $P_{\text{H}}$  & 17,130  & 8.596 - 7.466  & 0.742  & 0.27  & 12.26  & 2.00  & 0     & 6.1   & 6070   \\
    model 3 $P_{\text{H}}$  & 11,938  & 8.603 - 7.466  & 0.760  & 0.77  & 11.57  & 2.20  & 0     & 5.3   & 5672   \\
    model 4 $P_{\text{H}}$  & 11,948  & 8.600 - 7.479  & 0.759  & 0.75  & 11.50  & 2.28  & 0     & 5.0   & 5556   \\ 
    \end{tabular}
    \begin{tablenotes}
        \item Key parameters for all evaluated models consisting of H-He, water, rock and potentially iron. The models are separated into three groups: The first group consists of models with a high central density (see figure \ref{fig:density_profiles}). The second group uses the slower rotation period of Uranus ($P_{\text{V}}$) to calculate its structure. The third group uses the faster rotation period ($P_{\text{H}}$) instead. $XY$, $Z_{\text{H$_2$O}}$, $Z_{\text{SiO$_2$}}$ and $Z_{\text{Fe}}$ are the total { bulk abundances} of H-He, water, rock and iron, respectively. Finally, w:r is the water-to-rock ratio. {  Note that compared to pure rock models (table \ref{tab:results_rock}) these models tend to have lower central temperatures and higher { bulk heavy-elements abundance}. }
    \end{tablenotes}
    \label{tab:results_Z_mixtures}
\end{threeparttable}
\end{table*}

\subsection{The effect of winds on the inferred composition} \label{subsec:Wind_effect}
In \cite{Benno2022} we showed that winds penetrating to 1,100 km in Uranus can affect Uranus' density. 
In this section we investigate how such winds affect the inferred planetary temperature profile and bulk composition.  We compare the inferred composition of 5 models that have no winds with 4 models that include the dynamics contribution (winds going down to 1,100 km). This subset does not cover the entire parameter space of solutions but is a fair representation of the various models. 
\begin{figure}
    \centering
\includegraphics[width = 0.24\textwidth]{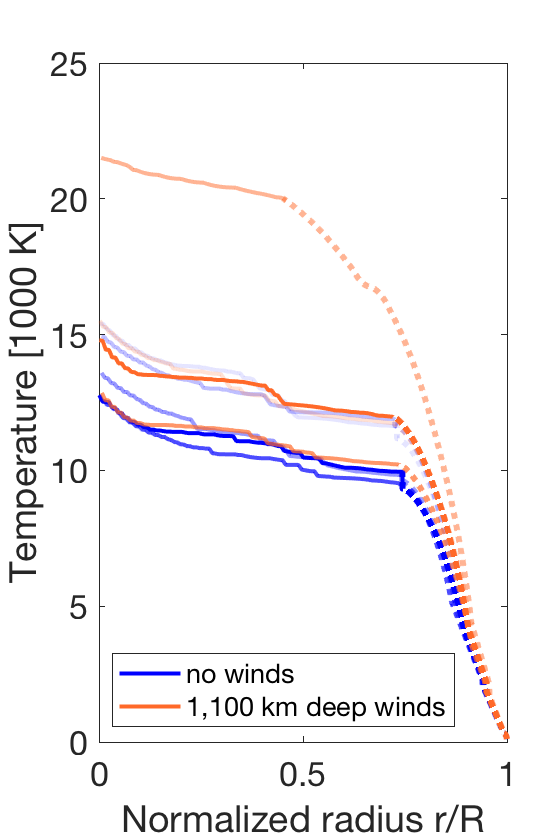}%
\includegraphics[width = 0.24\textwidth]{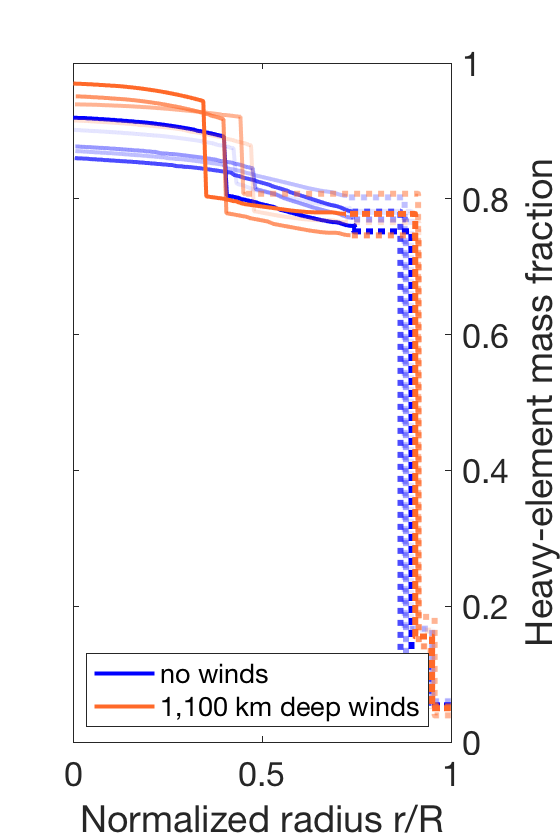}%
    \caption{Temperature (left panel) and metallicity (right panel) profiles of Uranus models with and without 1,100 km deep winds. Different orange and blue shades are used to make individual models traceable across the two panels. The dotted and solid part of each curve corresponds to  an { convective} and a non-{ convective} region, respectively.} 
    \label{fig:diff_winds_T_Z}
\end{figure} 
Figure \ref{fig:diff_winds_T_Z} shows the temperature (left panel) and metallicity (right panel) profiles of Uranus models with (green-themed) and without (blue-themed) winds. The figure suggests that both the temperature and the metallicity profile are not affected by the winds up to some minor shifts in the central region where $r_c \lesssim 0.4$. 
(incorporates roughly 1/4 of the planet's mass).
While we find that winds can lead to higher central metallicity, the { bulk heavy-element abundance} remains the same (around 11.5-12 $M_\oplus$ for both models, see table \ref{tab:results_rock}). 
This suggests that the higher metallicity in the core region of models with winds is at least partially compensated by a lower metallicity in the mantle (around r $\sim$ 0.4-0.8, incorporating  $\sim$ 2/3 of the planetary mass). Further, the higher central metallicities go along with a smaller core entropy. 
Finally, the transition radius between the { convective} and non-{ convective} regions is not affected in a systematic way. We conclude that winds going down to $\sim$ 1,100 km have only a minor effect on Uranus' inferred composition.  

\section{Summary and conclusion} \label{sec:summary_and_conclusion}
We present a new method to interpret the composition of Uranus from empirical models using a physical EoS. This method detects potential composition gradients and non-adiabatic regions. The resulting composition and temperature profiles are then fully consistent with the pressure-density profiles from the empirical models.  
We explore the thermal profile and composition distribution in Uranus using various density profiles and show that Uranus is expected to have a large non-{ convective} region. This non-{ convective} region can significantly affect the temperature profile and the distribution of heavy elements in Uranus. We next explore how the rotation period and the wind depth of the planet influence the inferred composition and water-to-rock ratio. A comparison between our findings and previous studies is also presented. 
Our key conclusions can be summarized as follows:

\begin{enumerate} [leftmargin=0.5cm,labelwidth=0pt]
    \item  {  The interpretation of the empirical structure models of Uranus presented by  \cite{Benno2022}, which are based on empirical density profiles, suggests that Uranus' interior  includes non-convective regions.  Uranus is found to have a convective atmosphere/mantle on top of a non-convective inner region which is stable. The transition between the convective and non-convective region is highly model-dependent and can vary between $\sim$40\% and $\sim$85\% of Uranus' radius.}
    \item {  The existence of non-adiabatic regions leads to much higher central temperatures in comparison to adiabatic models.  
    This leads to a higher inferred { bulk heavy-elements abundance} in Uranus (up to $\sim$10\%). }
    \item Different rotation periods affect the internal structure and composition of Uranus. A faster rotation period leads to hotter interiors (higher than 10,000 K, for $P_{\text{H}}$), a higher { bulk heavy-elements abundance} ($+\sim 0.5$ M$_\oplus$) and a lower water-to-rock ratio (5-6 compared to 7-15) and a larger { convective} region ($+\sim 35\%$). Therefore, an accurate determination of the rotation rate is required!
    \item Winds penetrating to $\sim 1000$ km have only a minor effect on Uranus' inferred { bulk heavy-element abundance}. However, they can change the heavy-element distribution and mildly affect the temperature profile.
    
\end{enumerate}

Our study clearly shows that the consideration of non-adiabatic regions in Uranus significantly affects its internal temperatures and the inferred planetary composition.  This in return has crucial implications for our understanding of Uranus' formation and evolution histories. 
The fact that in all the models we considered, a non-adiabatic interior has been inferred also has implications for exoplanet science. Since Uranus corresponds to an intermediate-mass planet, a planetary type that is very abundant in the galaxy, its characterization also reflects on exoplanets. In particular, our work suggests that non-adiabatic interiors must be included for exoplanetary characterization, and this can substantially affect the inferred composition of exoplanets. 
Finally, our study highlights the importance of a future mission to Uranus. We argue that measuring Uranus' rotation period is crucial since the uncertainty in the rotation period affects the temperature, composition, and structure of the planet. 
In addition, measuring Uranus' gravitational field with high accuracy would further constrain the possible density profiles and would put more strict limits on the depth of the winds. This would constrain Uranus' bulk composition and internal structure, and therefore improve our understanding of Uranus' formation and evolution. 

\begin{acknowledgements}
We thank Naor Movshovitz and Morris Podolak for many {fruitful} discussions. 
{ We also thank an anonymous referee for valuable comments that helped to improve the paper.}
Finally, we acknowledge support from the Swiss National Science Foundation (SNSF) under grant \texttt{\detokenize{200020_215634}}.
\end{acknowledgements}

%\clearpage
\bibliographystyle{aa}

\begin{appendix}

\section{Adiabatic vs non-adiabatic (pure rock)}  \label{app:adi_vs_nonadi_pure_rock}
\begin{figure}
    \centering
\includegraphics[width = 0.25\textwidth]{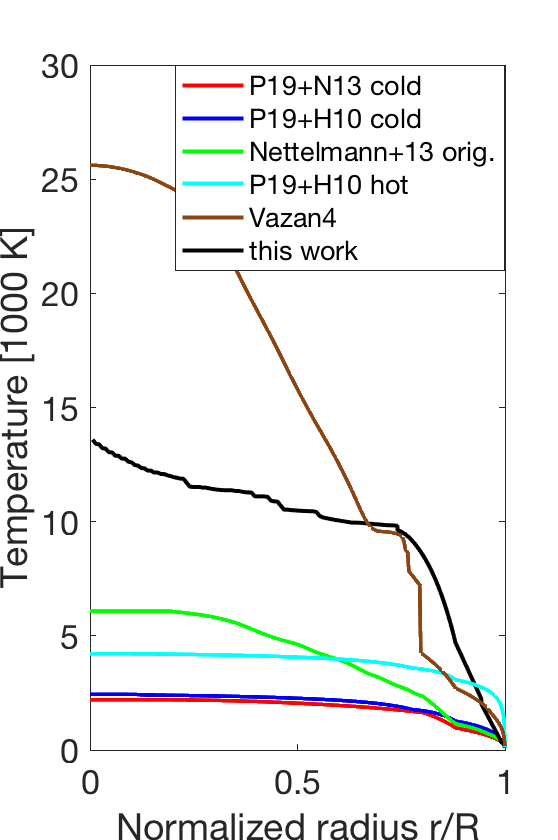}%
\includegraphics[width = 0.25\textwidth]{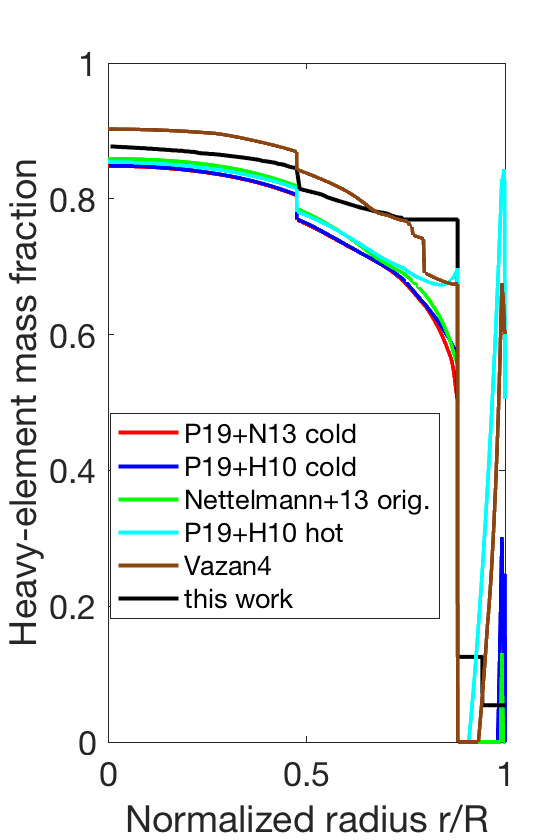}
    \caption{Temperature (left panel) and metallicity (right panel) of a single profile of Uranus evaluated with different temperature profiles. The heavy elements are represented by pure rock. "P19+N13 cold", "P19+H10 cold" and "Nettelmann+13 orig." are adiabatic temperature models, while "P19+H10 hot" and "Vazan4" are non-adiabatic temperature models.
}
    \label{fig:external_T_profiles}
\end{figure}

For a simpler comparison we analyze a model with only "rock" as heavy element.
Figure \ref{fig:external_T_profiles} shows the external temperature profiles (left panel) and the corresponding heavy elements distribution (right panel).  \\
It is clear from the figure that the non-adiabatic temperature profiles are hotter in comparison to the adiabatic ones. Therefore "P19+H10 hot" has a hotter interior than "P19+H10 cold". The same is true for the non-adiabatic solution of "Vazan4" or this work, both of which are crucially hotter than pure adiabatic solutions. 
This is expected as heat is transported  much slower in non-adiabatic regions which then stay hot for a longer time, leading to hotter interiors. 
\\
For each temperature, a composition profile is obtained (right panel of figure \ref{fig:external_T_profiles}. The outermost region (r>0.9) of the density profile clearly cannot be represented by any of the external temperature profiles. One reason could be that these external temperature profiles themselves are inferred by an assumed composition. Applying these temperature profiles on a different composition/density can cause unphysical behavior. This effect is even more pronounced in low temperature and pressure regimes such as the atmosphere. Below r$\sim$0.9, the pressure and temperature are large enough to diminish this behaviour. Note that for further analysis the uppermost $\sim10\%$ of the planet are neglected. This does not change the results to any significant extent, since the mass associate to these uppermost $\sim10\%$ is only about $\sim1\%$ of the planet's mass.  \\
It becomes clear from the figure, that small differences in the temperature profile do not affect both, the total amount or distribution of heavy elements in a significant way. In fact, the difference in the { bulk heavy-elements abundance} of "P19+N13 cold" (10.7 M$_\oplus$) and "Nettelmann+13 orig." (10.9 M$_\oplus$) is smaller than 2\%. This is expected as the applied equation of states are not strongly temperature dependent in large parts of the planet.
However, larger temperature differences as e.g., between the adiabatic "P19+N13 cold" and non-adiabatic "Vazan4" ({ bulk heavy-elements abundance} of 11.7 M$_\oplus$) affect the { bulk heavy-elements abundance} in the order of $\sim 9\%$. This result is to be expected, as in general the density of material decreases with increasing temperature. Therefore, a hot region with a given density can store more heavy elements than the same region with a cooler temperature.
We conclude that different adiabatic temperatures do not significantly alter the amount and distribution of heavy elements. However, the large temperature differences that may arise between adiabatic and non-adiabatic models are significant. We conclude that non-adiabatic temperature models change the planetary composition in a non-negligible way and that future studies have to account for the possible existence of non-convective regions. \\

\section{Comparison with published interior models}  \label{subsec:comp_with_published_sol}
In this section, we compare our models to several published adiabatic and non-adiabatic models of Uranus. 
Figure \ref{fig:density_profiles} already compares our density distributions with results from  \cite{Helled_2011_int_U_N,NETTELMANN2013, Vazan_2020_lumin_U}. Although we include different density profiles, some of the published models are not exactly covered with our sample. 
A potential explanation to that could be differences in the J$_2$ and J$_4$ values and their uncertainties. Uranus' gravitational harmonics were updated in 2014 \cite{Jacobson.2014} and, naturally, were not yet taken into account in the studies of \cite{Helled_2011_int_U_N, NETTELMANN2013}.  
Furthermore, the non-adiabatic solutions of \cite{Vazan_2020_lumin_U} are calculated by evolution models and therefore do not reproduce Uranus' gravity field. \\  
\begin{figure}
    \centering
\includegraphics[width = 0.5\textwidth]{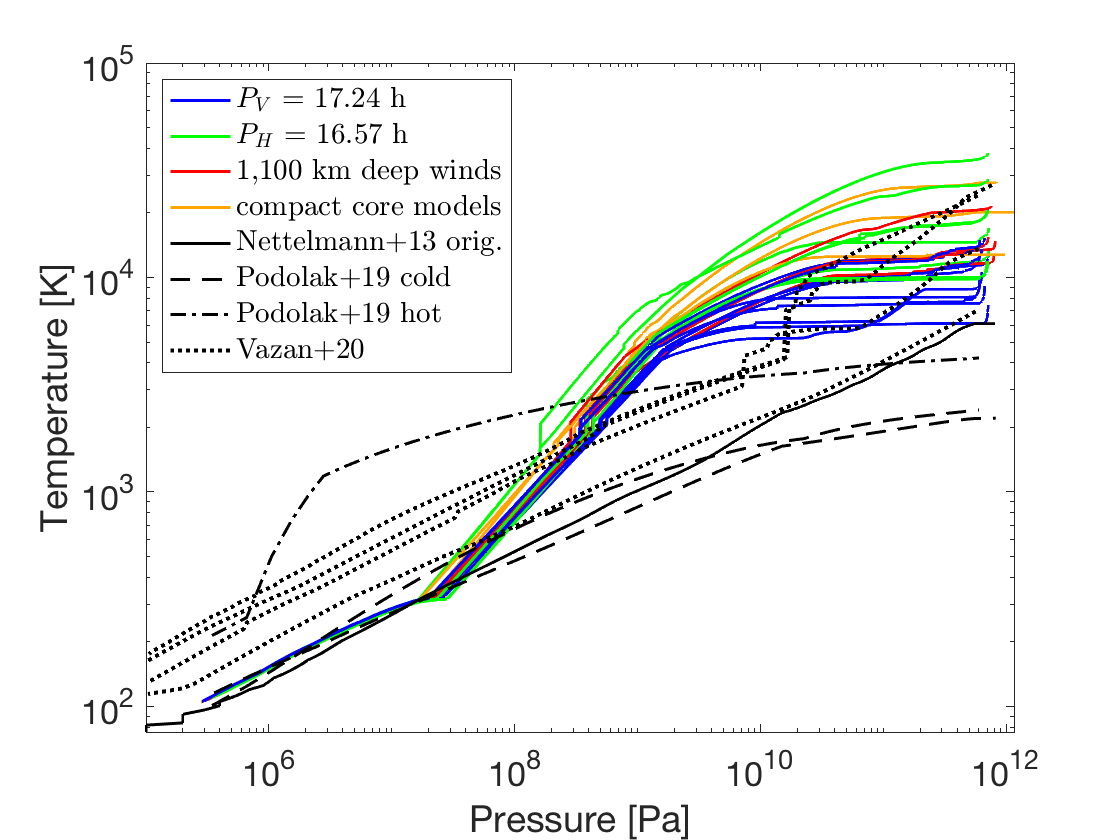}
    \caption{{  Temperature-Pressure dependence of all evaluated models (note color code).} For comparison, adiabatic and non-adiabatic solutions of published work is included. }
    \label{fig:logTlogP}
\end{figure}
Figure \ref{fig:logTlogP} collects all the temperature profiles of this work as well as published results including adiabatic and non-adiabatic models.  
It is clear from the figure that our models first follow very closely an adiabat (similar to \cite{NETTELMANN2013} and "Podolak+19 cold"). At $\sim$100 bar, the composition (and therefore also the adiabat) begins to change. This happens because, as discussed above, at this pressure the adiabatic atmosphere model from \cite{Hueso2020} ends and, from this point to the center of the planet, the density is described by the corresponding polytropic relation,  while at $\sim10^4$ bar, Uranus' interior clearly becomes non-adiabatic.
Note the flattening of the curves at very high pressures. 
At this depth, the temperature has little effect on the equations of state used, since they are mainly pressure-dependent and only slightly temperature-dependent.
It is clear that the non-adiabatic models lead to hotter deep interiors. This is crucial to consider in Uranus studies since it affects Uranus' cooling efficiency, and therefore its evolution history.

\section{Relation between radius and mass }
\begin{figure}
    \centering
\includegraphics[width = 0.5\textwidth]{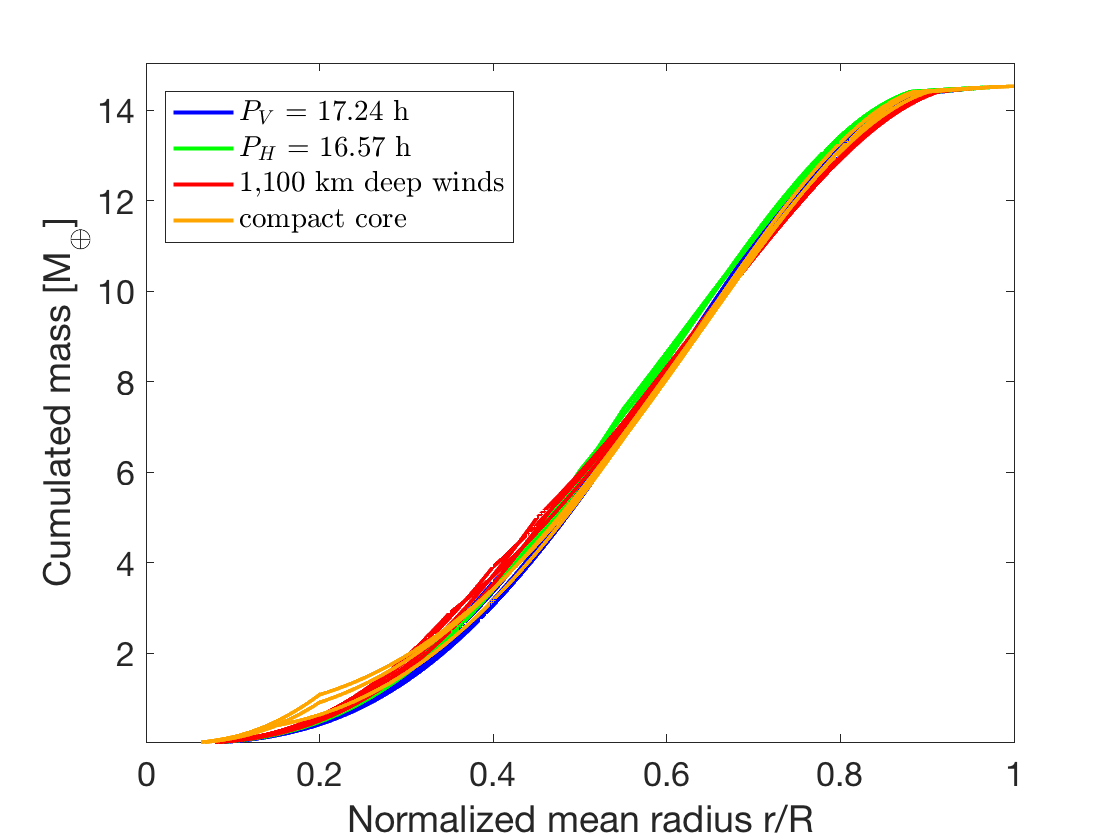}
    \caption{Relation between the accumulated planetary mass and normalized radius of the analyzed Uranus models. Note, small kinks may arise due to density discontinuities. }
    \label{fig:M_R_relation}
\end{figure}

There are at least two widely used approaches to present e.g., the density, pressure or temperature profile. The first is to plot the quantity against the planetary radius, the second way is to plot it against the planetary mass. Of course, both approaches have their justification as well as individual strengths and weaknesses. In the end, it may be the individual preference of each person which approach they choose. \\
In this paper we plot quantities such as the density, temperature, mass fractions usually against the normalized radius. However, for people that think more in terms of mass, we provide here the conversion of planetary radius and mass for a representative model. 
Figure \ref{fig:M_R_relation} illustrates the dependence between the planetary radius and accumulated mass of the analyzed models in this paper.

\section{{ Bulk heavy-element abundance} vs central temperature}
\begin{figure}
    \centering
\includegraphics[width = 0.5\textwidth]{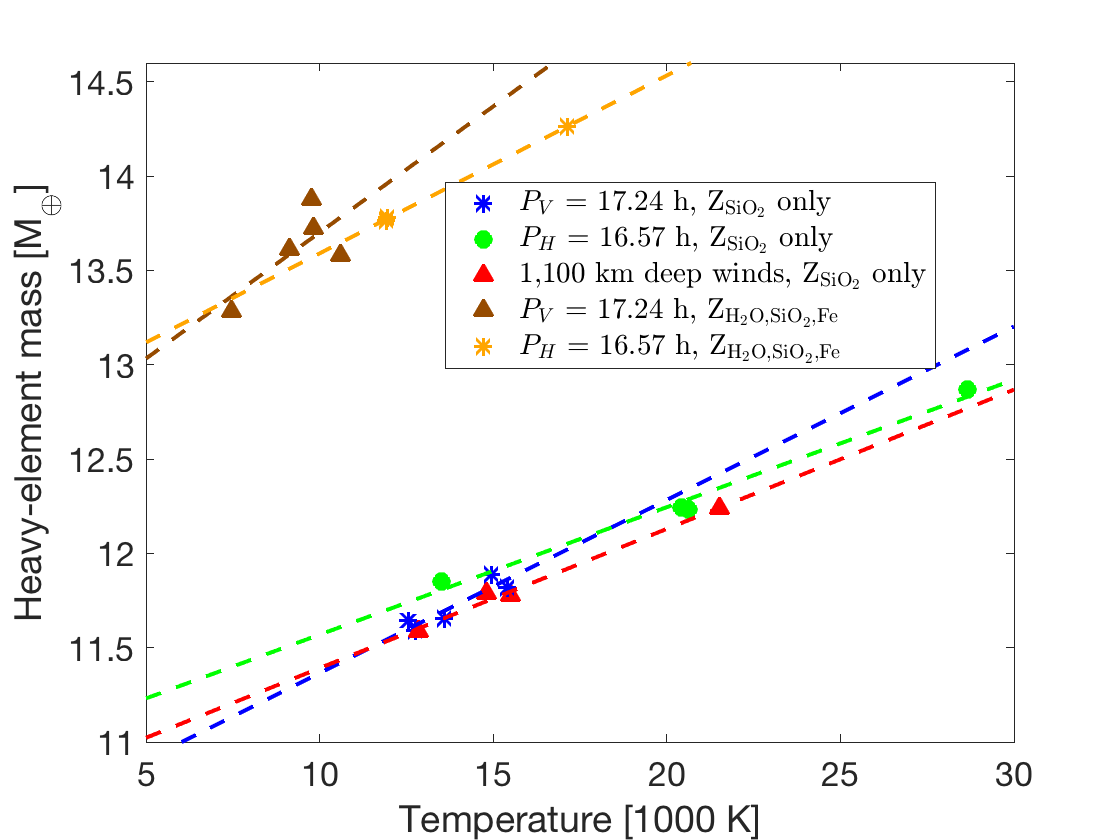}%
    \caption{How Uranus' total { bulk heavy-element abundance} depends on its central temperature. Each set of models is plotted separately (see legend). For each set, a dashed trend line is calculated and drawn in the corresponding color. 
    } 
    \label{fig:T_core_vs_metallicity}
\end{figure}

Figure \ref{fig:T_core_vs_metallicity} shows how Uranus' { bulk heavy-element abundance} depends on its central temperature. Models with different rotation periods and compositions are treated separately. Exact numbers depend on the chosen set of models, however, the general trend persists. Higher central temperatures lead to a higher { bulk heavy-element abundance} of Uranus. This finding is well known but the figure does quantify the effect and shows first-order trends.

\section{Supplementary Material}
\begin{figure*}
    \centering
\includegraphics[width = 0.5\textwidth]{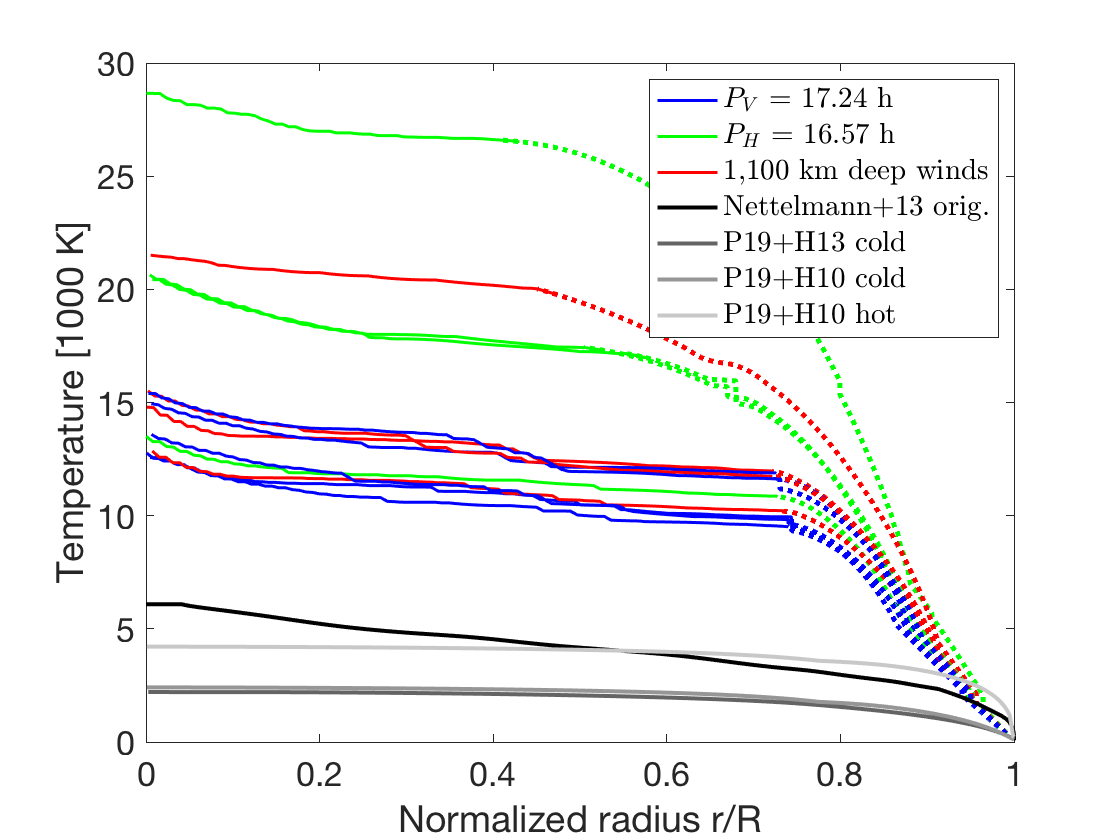}%
\includegraphics[width = 0.5\textwidth]{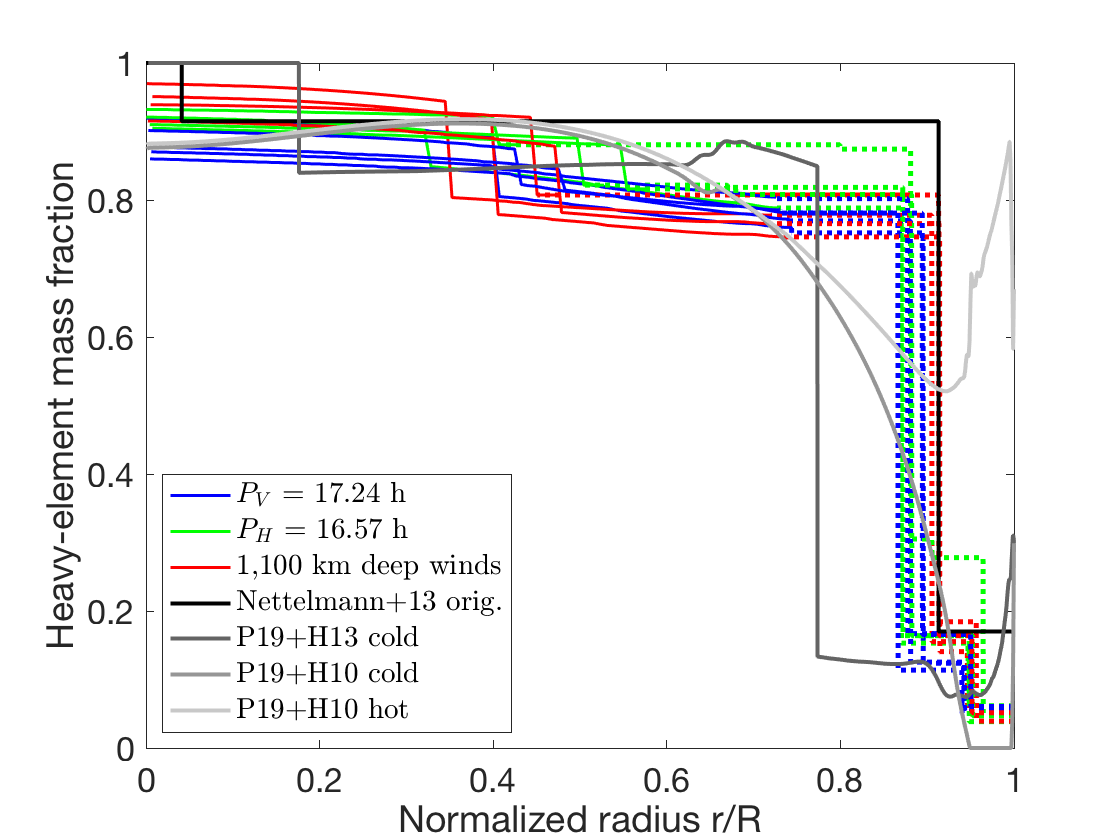}
    \caption{Comparison of our models with published results. The left (right) panel compares our models to the grey-shaded solutions of \protect\cite{NETTELMANN2013} and \protect\cite{Podolak_2019} in terms of temperature (metallicity). The { convective} and non-{ convective} regions of each model are indicated by dotted and solid parts in each curve, respectively.  All our models presented here consist of a proto-solar H-He mixture and pure rock as heavy elements. 
    }
    \label{fig:external_models_TZ}
\end{figure*}

\begin{figure}
    \centering
    \includegraphics[width = 0.5\textwidth]{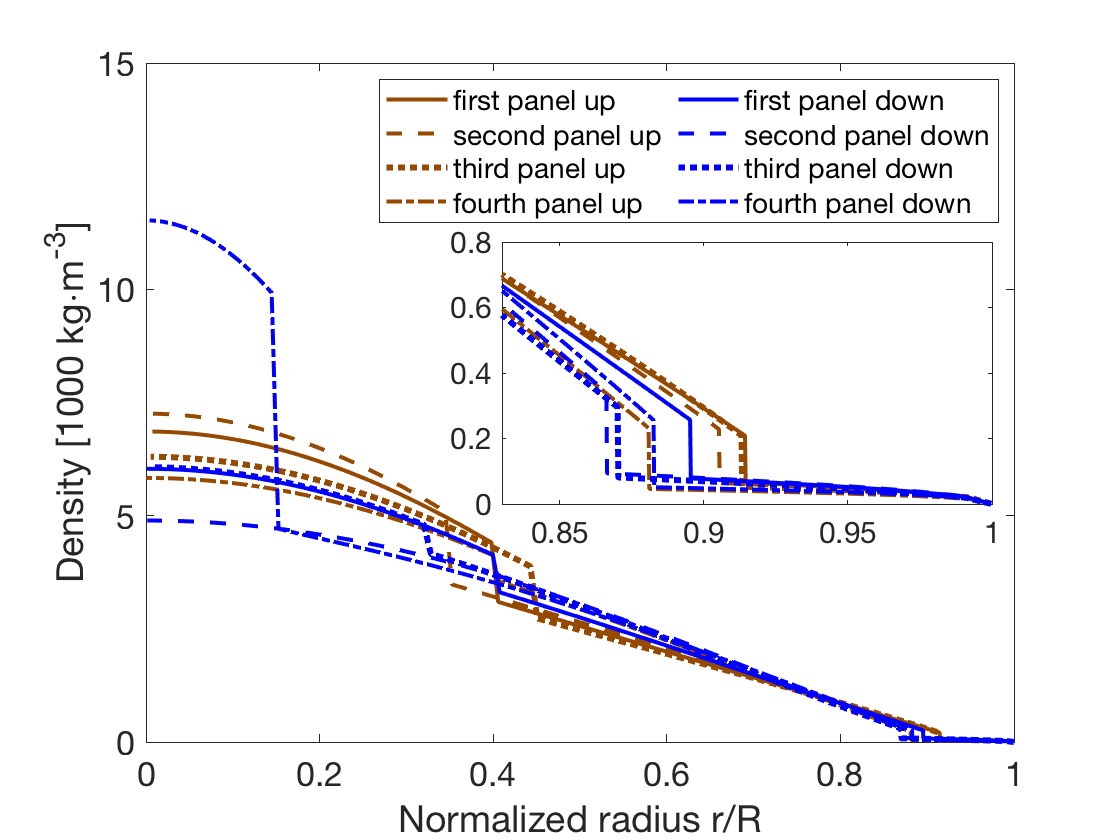}
    \caption{{Density profiles of each model shown in figure \ref{fig:T_rho_plots} }
    }
    \label{fig:density_figure_2}
\end{figure}

Figure \ref{fig:external_models_TZ} collects all models that consist of pure rock and compares them in terms of temperature (left panel) and metallicity (right panel) with published solutions. 
Figure \ref{fig:density_figure_2} shows for each panel in figure \ref{fig:T_rho_plots} its appropriate density profile. For better illustration, the transition region around $r\approx0.9$ is shown enlarged.

\end{appendix}

\end{document}